\documentclass[12pt]{iopart}
\usepackage{graphicx}
\usepackage{setstack}
\usepackage{amstext}
\usepackage{iopams} 
\usepackage{bbm}
\usepackage{bm}
\usepackage{xspace}
\usepackage{color}

\newcommand{\GG}{{\cal G}}
\newcommand{\mean}[1]{\left\langle #1 \right\rangle_{\epsilon}}

\def\bra#1{\mathinner{\langle{#1}|}}
\def\ket#1{\mathinner{|{#1}\rangle}}

\begin{document}

\title[Strong coupling expansion for the BH and JC lattice model]{Strong coupling expansion for the Bose-Hubbard and the Jaynes-Cummings lattice model}

\author{Christoph Heil, Wolfgang von der Linden}
\address{Institute of Theoretical and Computational Physics, Graz University of Technology, 8010 Graz, Austria}
\ead{cheil@sbox.tugraz.at}

\date{\today}

\begin{abstract}
A strong coupling expansion based on the Kato-Bloch perturbation theory, which has recently been proposed by Eckardt et al.~\cite{eckardt_process-chain_2009} and Teichmann et al.~\cite{teichmann_process-chain_2009}, is implemented in order to study various aspects of the Bose-Hubbard and the Jaynes-Cummings lattice model. The approach, which  allows to generate numerically all diagrams up to a desired  order in the interaction strength is generalized for disordered systems and for the Jaynes-Cummings lattice model.
Results for the Bose-Hubbard and the Jaynes-Cummings lattice model will be presented and compared with results from the variational cluster approach and density matrix renormalization group. Our focus will be on the Mott insulator to superfluid transition.
\end{abstract}


\maketitle

\section{\label{sec:introduction}Introduction}
The experimental progress in the field of ultracold atoms in periodic lattices~\cite{bloch_many-body_2008} allowed for a direct observation of quantum phase transitions. A very prominent example is the experiment by Greiner et al.~\cite{greiner_quantum_2002}, where the Mott insulator-superfluid~\cite{sachdev_quantum_2001} phase transition of ultracold Rubidium atoms trapped in a three-dimensional optical lattice has been observed. This phase transition is driven by two counteracting factors. On the one hand there is a repulsive interaction between particles on the same lattice site, trying to minimize the particle number fluctuation  per site and on the other hand there is a gain of kinetic energy whenever a particle tunnels from one site to another.

Consider a condensate of ultracold atoms in a Mott insulator phase, i.e. every lattice site is occupied by the same number of particles. One can then drive a phase transition to the superfluid phase, where the wave functions of the particles are no longer localized, by gradually decreasing the intensity of the lattice forming laser beams.

Because of this tunability of the experimental system~\cite{jaksch_cold_1998} it provides a great testing ground for various quantum many body theories. Two such theoretical models are the Bose-Hubbard (BH) and the Jaynes-Cummings lattice (JCL) model. While the BH model considers the competing interaction of kinetic and potential energy of bosons explained above~\cite{fisher_boson_1989}, the JCL model takes this thought further and also factors in the interaction of bosons with a two level atomic system. It therefore constitutes a pure quantum transcription of the semi-classical Rabi model.

Strong coupling approaches have already been used to study the BH model in great detail~\cite{freericks_strong-coupling_1996-1,elstner_dynamics_1999,damski_mott_2006,mering_one_2008,krutitsky_ultracold_2008}. In this work we treated BH and JC systems with a diagrammatic approach to the perturbation theory based on the Kato-Bloch expansion as suggested by Eckhard et al.~\cite{eckardt_process-chain_2009} and already applied to the ordered BH model by Teichmann~et~al.~\cite{teichmann_process-chain_2009}.

We extended the strong coupling Kato-Bloch approach to deal with disordered BH systems and the JCL model with additional degrees of freedom per unit cell.
In addition, we give details for an efficient numerical computation of the strong coupling diagrams and also outline the 
limitations and possible pitfalls of this approach.\\
This paper is organized as follows: The first part deals with the BH model. In \sref{sec:model} we are going to introduce the BH model in more mathematical detail. \Sref{subsec:energy_corrections} then goes on to explain how the Kato formalism can be incorporated to calculate corrections to the ground state energy. After this introduction we are going to discuss how this approach can be turned into numerical algorithms in \sref{subsec:algorithm}.\\
\Sref{subsec:bh_results} then shows a comparison of results calculated with the strong coupling approach with data obtained with the variational cluster approach (VCA)~\cite{knap_benchmarking_2010}. In \sref{subsec:mott_insulator_superfluid_phase_transition} we are going to discuss the Mott insulator-superfluid phase transition, how the phase boundary can be detected and which changes have to be applied to the present algorithms. Sec.~\ref{subsec:bh_ms_results} contains the results we obtained and also a discussion of the problems one encounters when trying to determine the phase boundary for a one-dimensional system within this approach. In \sref{subsec:1_dimensional_systems} we offer a workaround for this problem and show a comparison of our data with results obtained with a density matrix renormalization group (DMRG) approach by K\"uhner et al.~\cite{kuehner_one-dimensional_2000}. \Sref{subsec:disorder} explains how disordered BH systems can be treated with Kato-Bloch approach.\\
In the second part of this work we are going to deal with the JCL model. In \sref{sec:jaynes_cummings_model} we will offer a very short introduction to the JC-model and \sref{subsec:changes_algorithms_jc} then outlines the numerical transcription and what has to be changed in comparison to the BH implementation. The obtained results are shown in \sref{subsec:jc_energy_results}, where we compare the energies of various systems for different occupation numbers and dimensions. Additionally we compare data from the Kato-Bloch approach with results obtained with VCA~\cite{knap_polaritonic_2011}. The calculation of the Mott insulator-superfluid phase transition is explained in \sref{subsec:jc_mott_insulator_superfluid_transition}.\\
We conclude this work in \sref{sec:conclusion} with a summary of the obtained results, insights and a short outlook. 

\section{\label{sec:model}The Bose-Hubbard model}
The Bose-Hubbard model is used to describe bosonic particles in a lattice at very low temperatures~\cite{fisher_boson_1989-1} and has been treated with many different methods, including the mean-field method~\cite{fisher_boson_1989}, strong-coupling approaches~\cite{freericks_strong-coupling_1996-1,elstner_dynamics_1999,teichmann_process-chain_2009}, DMRG~\cite{kuehner_phases_1998,rapsch_density_1999,kollath_spatial_2004,kollath_quench_2007,ejima_dynamic_2011} and quantum monte carlo (QMC) methods~\cite{batrouni_world-line_1992,wessel_quantum_2004,capogrosso-sansone_phase_2007,capogrosso-sansone_monte_2008}. 
A particularly efficient strong coupling approach, based on the Kato-Bloch perturbation theory, has been used by Teichmann et al.~\cite{teichmann_process-chain_2009}.
In the following we will present in short all the definitions necessary to understand the subsequent equations. For a more detailed description the reader is referred to the above mentioned works.

The BH hamiltonian $\hat{H}$, which we split into two terms suitable for strong coupling perturbation theory, reads
\begin{eqnarray}\label{eq:operators}
\hat{H} &= \hat H_{0}	+ \hat H_{1}	\\
\hat{H}_{0} &:= \frac{U}{2} \sum_i \hat{n}_i (\hat{n}_i-1) - \mu \sum_i \hat{n}_i - \sum_i \varepsilon_i \hat{n}_i\label{eq:operators2}\\
\hat{H}_{1} &:=t\;\sum_{b} \hat h_{b} \label{eq:operators3} \;.
\end{eqnarray}

$\hat H_{0}$ describs the atomic limit, while $\hat H_{1}$ covers the hopping processes.
$U$ stands for the strength of the local Coulomb interaction, $\mu$ represents the chemical potential in the grand canonical ensemble, $\varepsilon_i$ gives the strength of the disorder on site $i$, $\hat{n}$ is the particle number operator and $\hat{a}$ and $\hat{a}^\dagger$ are the well known bosonic annihilation and creation operators. The indices label the sites of the lattice. The sums in $\hat H_{0}$ run over all lattice sites, while the sum in $\hat H_{1}$ runs over all directed bonds (dibonds) $\hat h_{b}$ of nearest neighbour sites, i.e.:
\begin{eqnarray}
\hat h_{b} = a^{\dagger}_{b_{2}} a^{\phantom{\dagger}}_{b_{1}},\\
\text{with } \; b \in {\cal P} := \big\{(i_{2},i_{1})| i_{1/2}\in \{1,2,\ldots,N\}\big\}\;. \nonumber
\end{eqnarray}

To begin with, we consider a homogeneous system $\varepsilon_{i} =0$. Disorder can easily be accounted for in a second step as shown in \sref{subsec:disorder}.
Energies will be expressed in units of $U$.
The eigenvectors of $\hat H_{0}$ are occupation number tensor products
\begin{equation*}
|\mathbf{n}\rangle := \bigotimes_{i=1}^{N} |n_{i}\rangle_{i} \;,
\end{equation*}
where $N$ stands for the total number of sites and the number of bosons at site $i$ is given by the integer $n_{i}$.
The corresponding eigenvalues are
 \begin{eqnarray}\label{eq:E0}
\epsilon(\mathbf{n}) &:= \sum_{i=1}^{N} n_{i}\bigg(
\frac{1}{2}(n_{i}-1) - \mu
\bigg) 
\end{eqnarray}
The ground state of $\hat{H}_0$ is denoted by $|\mathbf{g}\rangle = \otimes_{i} |g\rangle_{i}$, where each site is occupied by the same number of bosons, $g$, which is determined by minimizing $\epsilon(\mathbf{n})$ for a given value of the chemical potential, leading to a ground state energy $\epsilon_0=\langle \mathbf{g}|\hat{H}_0|\mathbf{g}\rangle$. 
\section{\label{subsec:energy_corrections}Ground State Energy}
In order to calculate the ground state energy in strong coupling perturbation theory, we make use of the Kato-Bloch formalism~\cite{bloch_many-body_2008,kato_convergence_1950,gelfand_perturbation_1990,messiah_quantum_1999} , which yields closed expressions for every order of the perturbation, in contrast to standard  Schr\"odinger-Rayleigh perturbation theory. 
The strong coupling Kato-Bloch approach is described in great detail in refs.~\cite{eckardt_process-chain_2009,teichmann_process-chain_2009}.

Here we summarize the key points of the formalism necessary to understand the generalization towards disordered systems and the JCL hamiltonian. In addition we will present a different perspective that allows 
to exploit graph theoretical techniques in order to speed up the algorithmic implementation. In the form presented in the current section, the Kato-Bloch perturbation theory is only applicable to a non-degenerate ground state, but we will lift that constraint in \sref{subsec:1_dimensional_systems}.\\
The $n^{\text{th}}$ order correction for the ground state can be written as 
\begin{eqnarray}
\label{eq:energy_corr1}
     \Delta E^{(n)}_g &= \sum_{\{k_{\nu}\}^{(n-1)}}\;\bra{\mathbf{g}} H_{1}  S_{k_{n-1}} \dots H_{1} S_{k_{2}} H_{1} S_{k_{1}} H_{1} \ket{\mathbf{g}},
\end{eqnarray}   
where the sum runs over all sets
\hbox{
$\{k_{\nu }\}^{(n-1)} := \big\{ (k_{1},k_{2},\ldots,k_{n-1})\big\}$} of 
non-negative integers  $k_{\nu}$, which satisfy the following conditions~\cite{messiah_quantum_1999}
\begin{eqnarray}\label{eq:constrain}
     \sum_{l=1}^s k_l &\le s\;;\qquad\text{for } s=1,\ldots,n-2 \\
    \sum_{l=1}^{n-1} k_l &= n-1 \;.
\end{eqnarray}
The operators $S_{k}$ are diagonal in the occupation-number basis
\begin{eqnarray}
\bra{\mathbf{n}} S_{k} \ket{\mathbf{n'}} &= \delta_{\mathbf{n},\mathbf{n'}}\;S_{k}(\mathbf{n})\\
      S_k(\mathbf{n}) &= \cases{
 -\delta_{\mathbf{n},\mathbf{g}} \hspace{3cm} \text{for } k=0 \vspace{0.5cm} \\ 
    \frac{1-\delta_{\mathbf{n},\mathbf{g}}}{\left( \epsilon_{0} - \epsilon(\mathbf{n}) \right)^k} \hspace{1.8cm} \text{otherwise}} \label{eq:def_S.b}
\end{eqnarray}
Eventually, the total energy correction can be expressed as a series in powers of the perturbation strength~$t$ with expansion coefficients $\alpha^{2n}$
\begin{equation}
\Delta E_g = \sum\limits_n \alpha^{(2n)}(g) \; t^n \label{eq:kato_energy_power_series}\;.
\end{equation}
\subsection{Graphical representation}
\label{subsubsec:graphical_representation}
Next we will map the evaluation of \eref{eq:energy_corr1} onto a graphical problem. 
For simplicity we consider simple cubic lattices in $d$ dimensions. The generalization to other lattices is fairly straight forward.
We start out from a finite cubic lattice with $N$ lattice sites
enumerated in some suitable way and periodic boundary conditions.
The energy per site is computed for the thermodynamic limit, $N\to \infty$.
We consider \eref{eq:energy_corr1} for a specific order $n$ and sequence
$\{k_{\nu}\}^{(n-1)}$. From the $n$ factors $H_{1}$ we obtain a multiple sum over dibonds
\begin{eqnarray}
\bra{\mathbf{g}} H_{1}  S_{k_{n-1}} \dots H_{1} S_{k_{1}} H_{1} \ket{\mathbf{g}} \nonumber \\ 
= t^{n} \sum_{b^{(1)},\ldots,b^{(n)}}
\bra{\mathbf{g}} \hat h_{b^{(n)}}  S_{k_{n-1}} \dots \hat h_{b^{(2)}} S_{k_{1}} \hat h_{b^{(1)}} \ket{\mathbf{g}}\;.
\label{eq:aux1}
\end{eqnarray}

There are two important aspects worth noting. Firstly, the application of any of the dibond hopping operators $h_{b}$ on an occupation-number basis vector results in just another  such basis vector
\begin{equation*}
h_{b}\ket{\mathbf{n}} = \ket{\mathbf{n'}}\;.
\end{equation*}
Secondly, as the operators $S_{k}$ are diagonal in the occupation number basis they merely introduce weight factors. 
Hence, the index $\tau$ of the dibond operators $h_{b^{(\tau)}}$ defines an auxiliary time 
and $b^{(\tau)}$  can be any of the $2 d$ dibonds on the d-dimensional simple cubic lattice.
Each of the individual dibond hopping processes  can be depicted  as a directed line (vector) on the underlying lattice, connecting neighbouring sites (see for example \fref{fig:topology4_2}).
The resulting pattern of arrows can be considered as a labelled 
digraph~\cite{gibbons_algorithmic_1985}.

\begin{figure}
  \begin{center}
    \begin{tabular}{ccc}
     \includegraphics[width=0.15\linewidth]{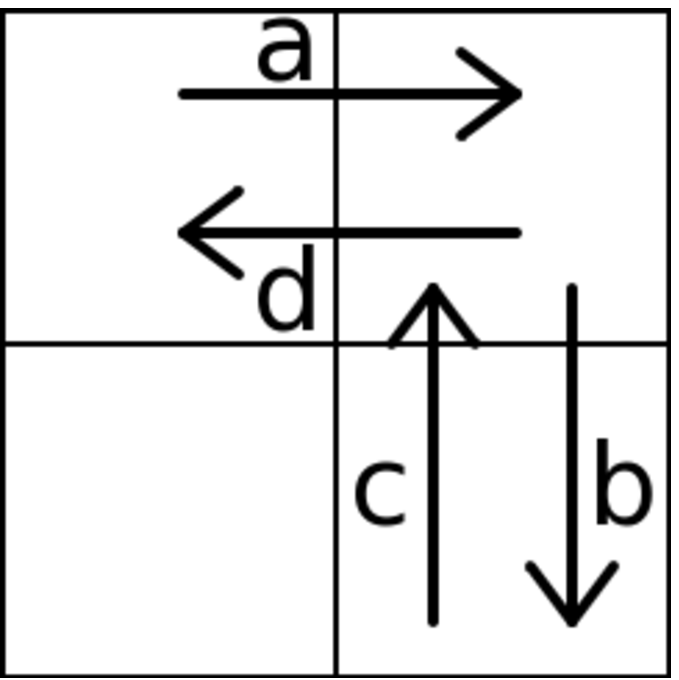} & \hspace{0.85cm} &
     \includegraphics[width=0.15\linewidth]{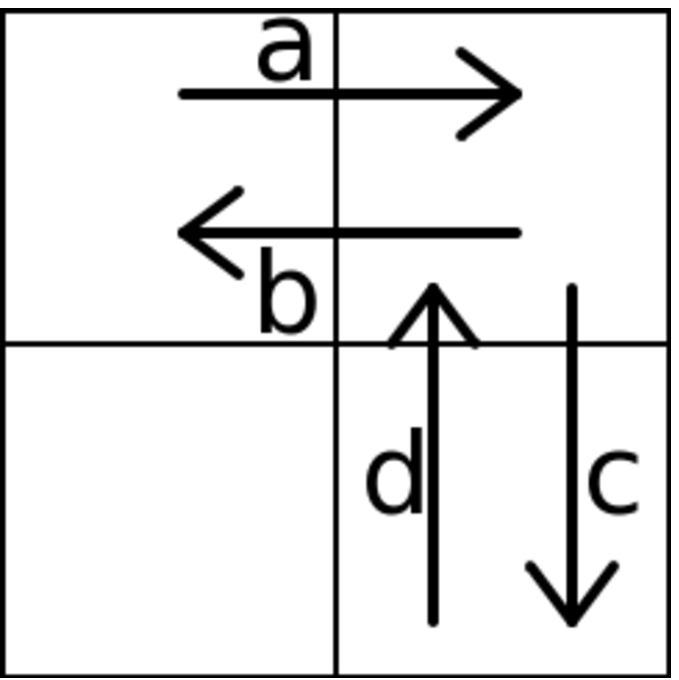} \\
      (a) &  & (b)
    \end{tabular}
    \caption{Two possible time resolved dibond sequences of a digraph of $4^{\text{th}}$ order.}
    \label{fig:path4perm}
  \end{center}
\end{figure}

So far, the sum over dibonds in \eref{eq:aux1} can be replaced by a sum over all labeled digraphs $\GG$, consisting of directed nearest neighbour lines. 
Such a digraph covers a certain set of lattice sites, which in ordered form be
\begin{equation*}
i_{1}< i_{2} < \ldots < i_{m}\;,
\end{equation*}
where $m$ is the number of vertices of the digraph.
Due to the homogeneity of the lattice we can renumber the lattice sites in sequential order  without changing the resulting matrix elements in \eref{eq:aux1}.
\begin{equation*}
i_{\nu} \to \nu\;;\qquad \nu=1,\ldots,m\;.
\end{equation*}
The digraph defines a set ${\cal B}=\{b^{(i)}\}$ of dibonds, but not the order in which they occur in \eref{eq:aux1}. 
Therefore, in addition to the sum over all digraphs $\GG$ with $n$ directed lines, 
we have to sum over all distinct sequences of dibond operators occurring in $\GG$. Two such distinct sequences of the same digraph are depicted in \fref{fig:path4perm}. Obviously, multiple dibonds are indistinguishable and need not be permuted.

For each dibond sequence $\vec b$, we compute the corresponding matrix elements (weights of the digraph) 
\begin{eqnarray}
w(\GG)&= \bra{\mathbf{g}} \hat h_{b^{(n)}}  S_{k_{n-1}} \dots \hat h_{b^{(2)}} S_{k_{1}} \hat h_{b^{(1)}} \ket{\mathbf{g}}\nonumber\\
&= \bra{g_{1},\ldots,g_{m}} \hat h_{b^{(n)}}  S_{k_{n-1}} \dots \hat h_{b^{(2)}} S_{k_{1}} \hat h_{b^{(1)}} \ket{g_{1},\ldots,g_{m}}\;,
\label{eq:graph.weight}
\end{eqnarray}
where, due to the tensor structure of the vectors and the structure of the operators, only the sites $1,\ldots,m$ are involved.
An immediate consequence is that the weight of all isomorphic digraphs 
is the same. Since the computation of the matrix elements is the most 
cpu-time consuming step, it is expedient to represent the digraphs by topologically different members and the corresponding multiplicities.
\subsubsection{Translational invariance}
One class of isomorphism is due to translational invariance.
Each digraph, apart from the lattice-site labels, occurs in $N$ copies on the lattice with periodic boundary conditions, which results in an overall factor $N$.
In order to avoid the computation of $N$ identical contributions, we 
compute the energy per site, by restricting the sum over all digraphs to those attached to the origin. Graphs attached 
to the origin shall be defined by the condition
\begin{equation}\label{eq:shift.to.origin}
\min_{i} \vec x^{(i)} =\vec 0\;,
\end{equation}
where $\vec x^{(i)}$ is  the position  of the $i$-th
site on the underlying lattice, covered by the digraph.
\subsubsection{Additional constraints on the graphs}
Yet not all digraphs contribute to the energy. There are two constraints
for non-zero contributions.
Firstly, the number of dibonds entering a site has to balance the number of 
dibonds leaving it ({\it local particle conservation rule}). 
In graph-theory language the digraph is said to be balanced and consequently it has no sources or sinks, i.e. sites with only one dibond attached to it. 

A second constraint is due to the linked cluster theorem.
As shown in ref.~\cite{eckardt_process-chain_2009,teichmann_process-chain_2009} and references therein, contributions of disconnected graphs cancel. A simple argument is given by the extensivity of the energy which has to be proportional to $N$. Disconnected graphs yield higher orders in $N$, since each disconnected part can be placed roughly speaking on $N$ different sites.

In summary, admissible digraphs are connected and balanced. From graph theory it is known~\cite{gibbons_algorithmic_1985} that such graphs are Eulerian, i.e. they can be traversed along the directed lines by visiting each directed line exactly once. This very feature will be used to generate all digraphs. Examples of several low-order diagrams are given in 
refs.~\cite{eckardt_process-chain_2009,teichmann_process-chain_2009}.
\subsection{Algorithm}
\label{subsec:algorithm}
Here we give an algorithm to generate the topologically different, connected and balanced digraphs on a simple cubic lattice.
\subsubsection{Generation of topologically different digraphs}
We start at the origin of the lattice and generate an Eulerian circuit of length $n$ by tracing a tour sequentially to neighbouring sites.
At every site there are $2d$ choices to proceed. The problem is equivalent to searching in trees and we  applied a depth-first search algorithm up to depth $n$. If we return to the origin after these $n$ steps, the corresponding path is shifted such that it fulfills  \eref{eq:shift.to.origin} and is added to a list. Multiple entries of the same path are discarded. The same path can occur several times, as we can consider  any site of the path 
as origin, from which the path is initiated.
This multiplicity is an artifact of the algorithm and has to be eliminated.

Next we need to identify topologically different digraphs.
Digraphs are isomorphic if and only if for some ordering of their vertices, their adjacency matrices~\cite{brualdi_combinatorial_2008} are equal.

Let $\{i_{1},i_{2},\ldots,i_{n+1}\}$ be the sequence of site-indices encountered during the tour. The adjacency matrix, which includes the information which sites of a graph are connected with each other, can be computed in the following way
\begin{equation*}
M_{i,j} =\sum_{l=1}^{n}   \delta_{i,i_{l}} \delta_{j,i_{l+1}}\;,
\end{equation*}
where $M_{i,j}$ is the number of times the path goes from site $i$ to site $j$
or in other words, the number of dibonds from site $i$ to site $j$.

Two digraphs are isomorphic if and only if  the corresponding matrices are similar 
\begin{equation*}
M' = P M P^{-1}\;.
\end{equation*}
with $P$ being a permutation matrix~\cite{brualdi_combinatorial_2008}. Provided two graphs are isomorphic, trace and determinant of $M^{\nu}$
and  $M'^{\nu}$ for any integer power $\nu$ have to agree. This can be used as a quick tests to rule out isomorphism.

\begin{figure}
  \begin{center}
     \includegraphics[width=0.3\linewidth]{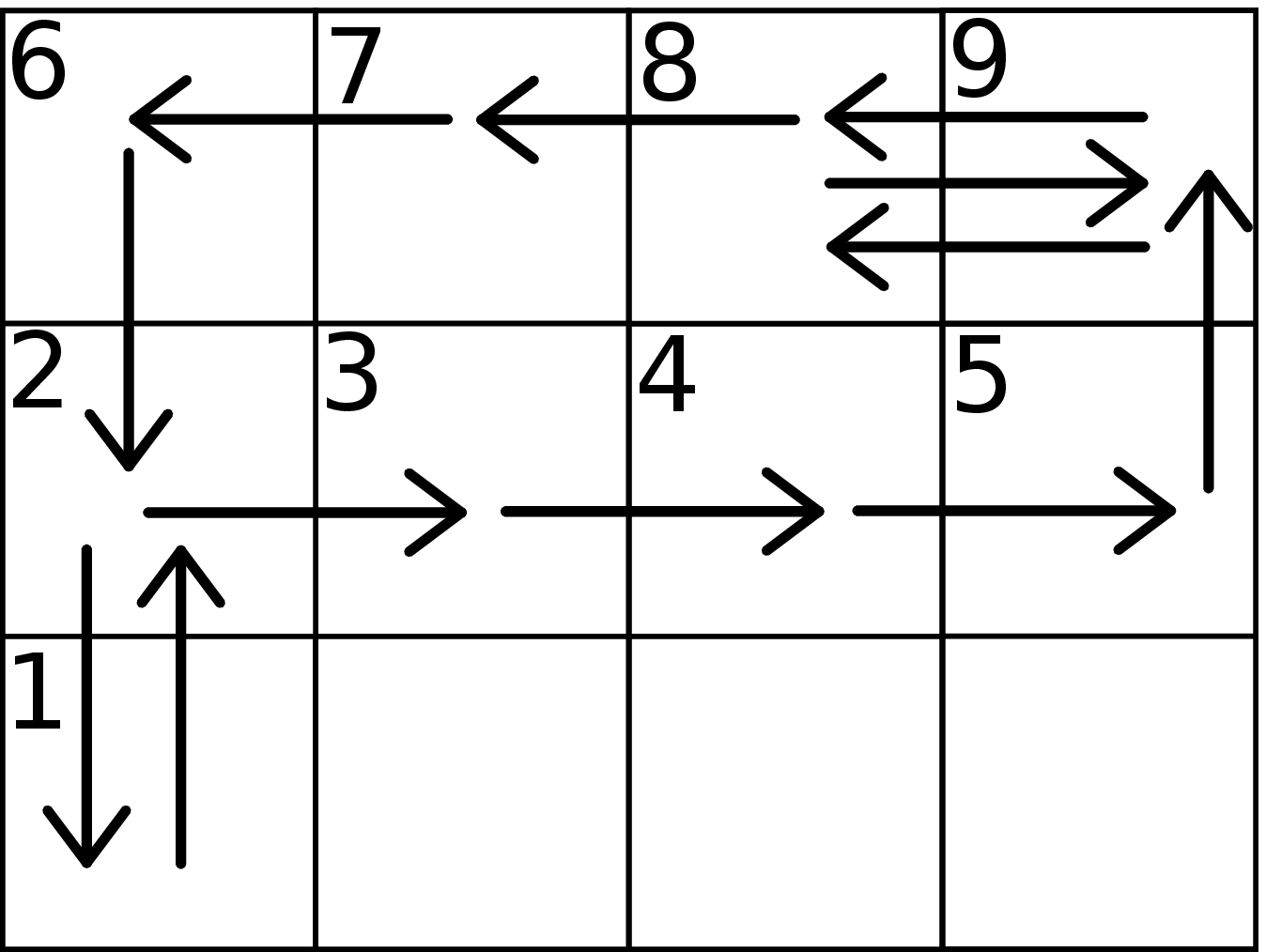}
	\hspace{0.5cm}
     \includegraphics[width=0.3\linewidth]{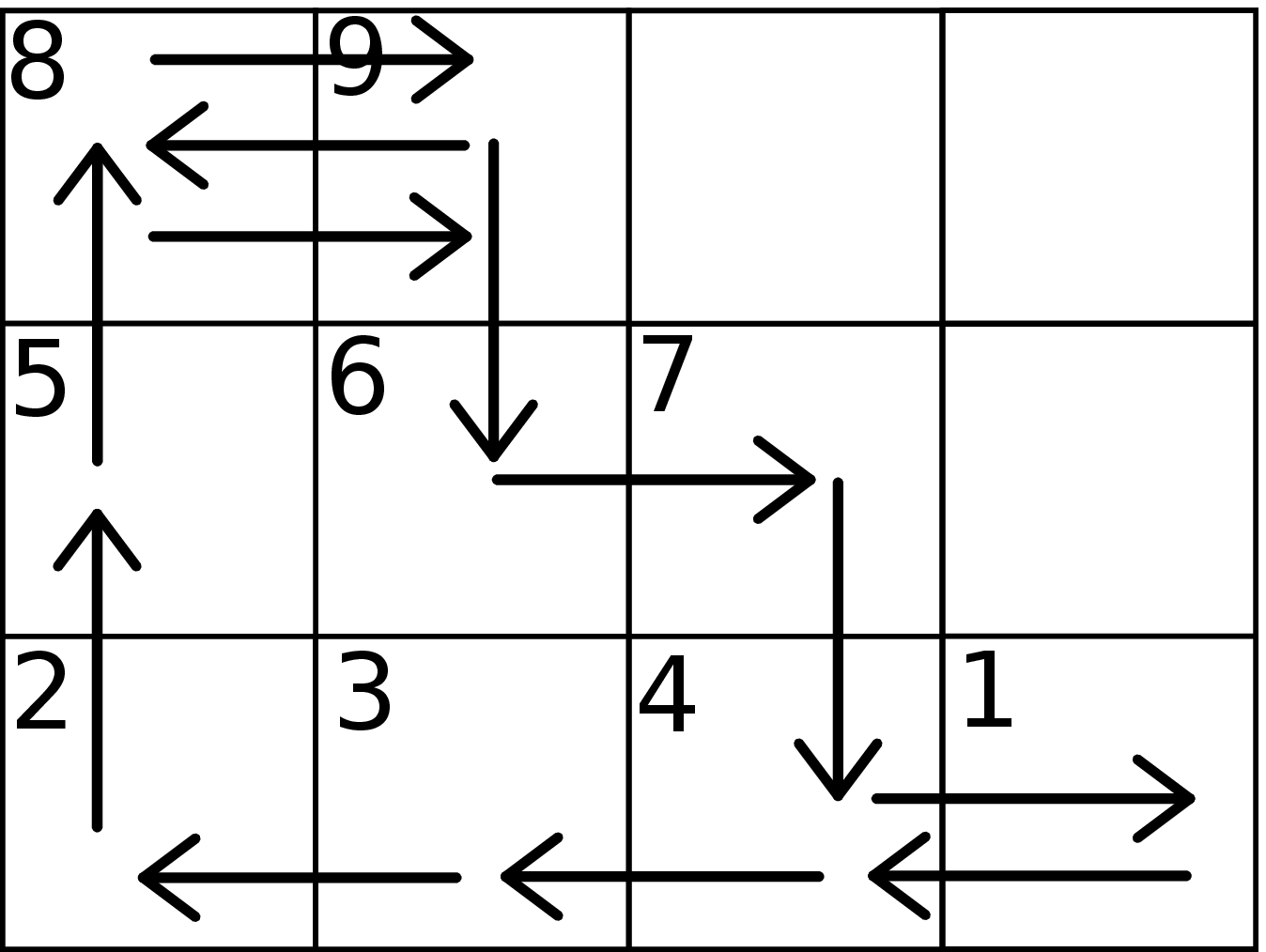} 
    \caption{Two identical digraphs of order $n=12$. The generating paths are attached to the origin.}
    \label{fig:topology4_2}
  \end{center}
\end{figure}

A representative example of two digraphs that pass all these tests is depicted in~\fref{fig:topology4_2}. 
The corresponding adjacency matrix of the left digraph is
\begin{equation}
M_1 = 
\left(\begin{array}{ccccccccc}
0 & 1 & 0 & 0 & 0 & 0 & 0 & 0 & 0\\ 
1 & 0 & 1 & 0 & 0 & 0 & 0 & 0 & 0\\ 
0 & 0 & 0 & 1 & 0 & 0 & 0 & 0 & 0\\ 
0 & 0 & 0 & 0 & 1 & 0 & 0 & 0 & 0\\ 
0 & 0 & 0 & 0 & 0 & 0 & 0 & 0 & 1\\ 
0 & 1 & 0 & 0 & 0 & 0 & 0 & 0 & 0\\ 
0 & 0 & 0 & 0 & 0 & 1 & 0 & 0 & 0\\ 
0 & 0 & 0 & 0 & 0 & 0 & 1 & 0 & 1\\ 
0 & 0 & 0 & 0 & 0 & 0 & 0 & 2 & 0
\end{array}\right)
\end{equation}
and the adjacency matrix of the right digraph reads
\begin{equation}
M_2 = 
\left(\begin{array}{ccccccccc}
0 & 0 & 0 & 1 & 0 & 0 & 0 & 0 & 0\\ 
0 & 0 & 0 & 0 & 1 & 0 & 0 & 0 & 0\\ 
0 & 1 & 0 & 0 & 0 & 0 & 0 & 0 & 0\\ 
1 & 0 & 1 & 0 & 0 & 0 & 0 & 0 & 0\\ 
0 & 0 & 0 & 0 & 0 & 0 & 0 & 1 & 0\\ 
0 & 0 & 0 & 0 & 0 & 0 & 1 & 0 & 0\\ 
0 & 0 & 0 & 1 & 0 & 0 & 0 & 0 & 0\\ 
0 & 0 & 0 & 0 & 0 & 0 & 0 & 0 & 2\\ 
0 & 0 & 0 & 0 & 0 & 1 & 0 & 1 & 0
\end{array}\right)
\end{equation}
At first glance, the two matrices seem to be different, but the distinction stems from the labeling. 
If we denote by $P$ the matrix associated with the permutation
\begin{equation*}
\left((1),(2,4),(3),(5),(6,7),(8,9)\right)\;,
\end{equation*}
then we can easily verify the similarity of $M_{1}$ and $M_{2}$
\begin{equation*}
M_{2} = P^{-1} M_{1} P\;.
\end{equation*}
This algorithm then generates a list of unlabeled (topologically different) digraphs, along with their multiplicities.

\subsubsection{Computation of the energy correction}
In order to evaluate   
the energy correction per site of \eref{eq:energy_corr1} we have to go through  all digraphs in the list, compute their contribution to the matrix element and multiply it by the multiplicity.
The contribution of one digraph is determined in the following way.
The digraph $\GG$ defines a set of dibond operators. 
We attach auxiliary times $\tau = 1,\ldots,n$ to the dibond operators
in all possible but distinguishable ways. Each time sequence $(b^{(n)},\ldots,b^{(1)})$ defines an additive contribution to \eref{eq:energy_corr1}.\\
There is an implicit condition due to the fact that 
\begin{equation}
\bra{\mathbf{g}} h_{b} \ket{\mathbf{g}} = 0\,  \label{eq:bond_operator}
\end{equation}
for every  dibond operator. Consequently, $k_{1}\ne 0$ and
$k_{n-1}\ne 0$ and for the same reason, consecutive $k$-values cannot be zero simultaneously, i.e. $k_{i}\; k_{i+1}\ne 0$.

\subsubsection{Summary of the algorithm}
\label{subsubsec:evaluating_energy}

\begin{itemize}
\item {\it Generation of paths:}
First we generate all closed paths of a given order $n$, beginning at the origin. 
\item {\it Translational invariance:}
Each path is attached  to the origin as defined by \eref{eq:shift.to.origin}.
Multiple occurrences of the same path are eliminated. 
\item {\it Digraphs:}
The paths are translated into digraphs.
Site labels are omitted and (topologically) unique digraphs are 
stored along with their multiplicities.
\item {\it Dibond sequences:}
For each digraph, all distinct time dependent dibond sequences are generated.
\item {\it Kato-Bloch summation:}
For each dibond sequence, the compatible Kato-Bloch indices $\{k_{\nu}\}^{(n-1)}$ are determined and the contribution to the energy correction \eref{eq:energy_corr1} are computed. Care is taken when admissible subgraphs occur.
\end{itemize}

\subsection{Test result}
\label{subsec:bh_results}
\begin{figure}
  \centering
    \includegraphics[width=0.8\linewidth]{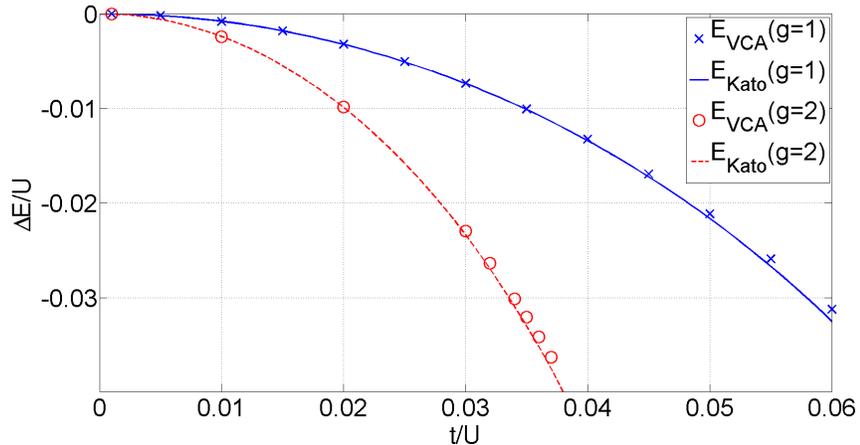}
      \caption{Energy corrections with the present approach (up to $8^{\text{th}}$ order) for 1 dimension and filling factors $g=1$ and $g=2$ compared with results obtained with VCA~\cite{knap_benchmarking_2010}.}
  \label{fig:energy_comp_knap}
\end{figure}

In order to test the numerical algorithm 
we compared the ground state energy as a function of the 
hopping strength for the one dimensional BH model with the
results obtained by VCA~\cite{knap_benchmarking_2010}. An example is given in \fref{fig:energy_comp_knap} and as one can see, the results are in very good agreement.
More results for the BH model obtained with the strong coupling Kato-Bloch approach can be found in ref.~\cite{teichmann_process-chain_2009}.
\section{Mott insulator-superfluid phase transition}
\label{subsec:mott_insulator_superfluid_phase_transition}
We will now address the phase boundary between the Mott insulator and the superfluid phase, which can easily be computed in the frame of the strong coupling Kato-Bloch formalism. 
There are essentially two approaches, for fixed $\mu$ or for fixed $t$.

The latter is driven by charge fluctuations and the phase boundary can be detected by computing the energy $\Delta E^{\pm1}_{g}$ it takes to add or subtract a 'defect' particle as has been shown in ref.~\cite{freericks_strong-coupling_1996-1} (see \sref{subsec:1_dimensional_systems}).

At fixed $\mu$ and density, the transition is driven by phase fluctuations. 
To detect this transition, we introduce an 'Effective Potential'~\cite{teichmann_process-chain_2009,negele_quantum_1988,dos_santos_quantum_2009}.
\subsection{Phase Boundary Criterion}\label{phase_boundary_criterion}
As a very detailed description of the method of the 'Effective Potential' is given in ref.~\cite{teichmann_process-chain_2009}, we will do without a derivation and just present the necessary equations and definitions.\\
In order to describe the superfluid phase, we have to add a source and a drain term to the BH hamiltonian
\begin{equation}
\widetilde{H} =\underbrace{\hat{H}_0 + \hat{H}_{hop}}_{H_\text{BH}} +\sum\limits_i \left( \eta^* \hat{a}_i + \eta \hat{a}_i^\dagger \right)~, \nonumber
\end{equation}
where  $\eta$ is the strength enforcing particle number fluctuations.
The order parameter $\psi=\langle \hat{a}_i\rangle_\eta$ can then be written as a power series in $\eta$
\begin{equation}\label{eq:kato6}
    \psi = (c_2  + 2 c_4 | \eta|^2 + O(| \eta|^4)) \;\eta\;.
\end{equation}
The coefficients $c_{2n}$ that appear in \eref{eq:kato6} are expanded in power series of the hopping parameter $t$:
\begin{equation}
	c_{2 l} = \sum\limits_{n=0}^\infty \alpha_{2 l}^{(n)} t^n \label{eq:free_energy_constants}
\end{equation}
As pointed out in ref.~\cite{teichmann_process-chain_2009}, $c_{2}$ can be considered as a susceptibility
\begin{equation*}
c_{2} = \left(\frac{\partial \Psi}{\partial  \eta}\right)\bigg|_{\eta\to 0 }\;,
\end{equation*}
of the system to develop an order parameter driven by the source-and-drain term of strength $\eta$. Hence, the Mott-to-superfluid phase transition occurs at those points in the $\mu-t-$ phase diagram, where $c_{2}$ diverges.

The convergence radius of the  power series of the function $\alpha_{2}^{(\nu)}(t)$ in 
\eref{eq:free_energy_constants} is according to  d'Alembert's law~\cite{whittaker_course_1927} given by
\begin{equation}\label{eq:convergence_radius}
	t^{*} = \lim\limits_{\nu \rightarrow \infty} \left| \frac{\alpha^{(\nu-1)}_{2}}{\alpha^{(\nu)}_{2}} \right| \;.
\end{equation}
Hence, for given chemical potential, $t^{*}$ marks the transition from the Mott insulator to the superfluid phase.

In order to use the strong coupling Kato-Bloch formalism the perturbation term has to be modified
\begin{equation*}
H'_{1} = H_{1} + \sum\limits_i \left( \eta^* \hat{a}_i + \eta \hat{a}_i^\dagger \right)\;.
\end{equation*}
In analogy to \eref{eq:energy_corr1} and \eref{eq:aux1}
we have
\begin{eqnarray}
\alpha_{2}^{(n-2)} = \sum_{\{k_{\nu}\}^{(n-1)}} \sum_{b^{(1)},\ldots,b^{(n)}}
\bra{\mathbf{g}} \hat o^{(n)}  S_{k_{n-1}} \dots \hat o^{(2)} S_{k_{1}} \hat o^{(1)} \ket{\mathbf{g}}\;.
\label{eq:aux11}
\end{eqnarray}
The operators $\hat o^{(\tau)}$ are either 
dibond operators $h_{b^{(\tau)}}$ as in \eref{eq:operators3} 
or 
 individual particle creation $\hat{a}_i^\dagger$  or annihilation  operators 
 $\hat{a}_i^{\phantom{\dagger}}$ at site $i$.
As a matter of fact,
since we are only interested in the term $c_{2}$,
exactly one creation and one annihilation operator have to be present.

\subsection{Graphical representation}

For the $n^{th}$ order contribution ($\alpha_2^{(n)}$) the graphical elements are $n$ nearest neighbour directed lines, one source term and one drain term. 
By virtue of the linked cluster theorem the graph has to be connected and the 
source and drain symbols are located on the vertices of the graph.
On top of that, particle conservation demands
\begin{equation*}
d^{+}(i) - d ^{-}(i) = N^{+}(i) - N^{-}(i)\;,
\end{equation*}
where $d^{+/-}(i)$ are out-/in-degrees of vertex $i$ and
$N^{+/-}(i)$ is the number of source-/drain-symbols of vertex $i$.
In other words, vertices with no source/drain or with a source-drain pair are balanced, while a vertex with only a source/drain symbol has to have one remaining, not compensated outgoing/incoming directed line.

\subsubsection{Construction of admissible digraphs}

\begin{figure}
  \begin{center}
    \begin{tabular}{ccccccc}
     \includegraphics[width=0.15\linewidth]{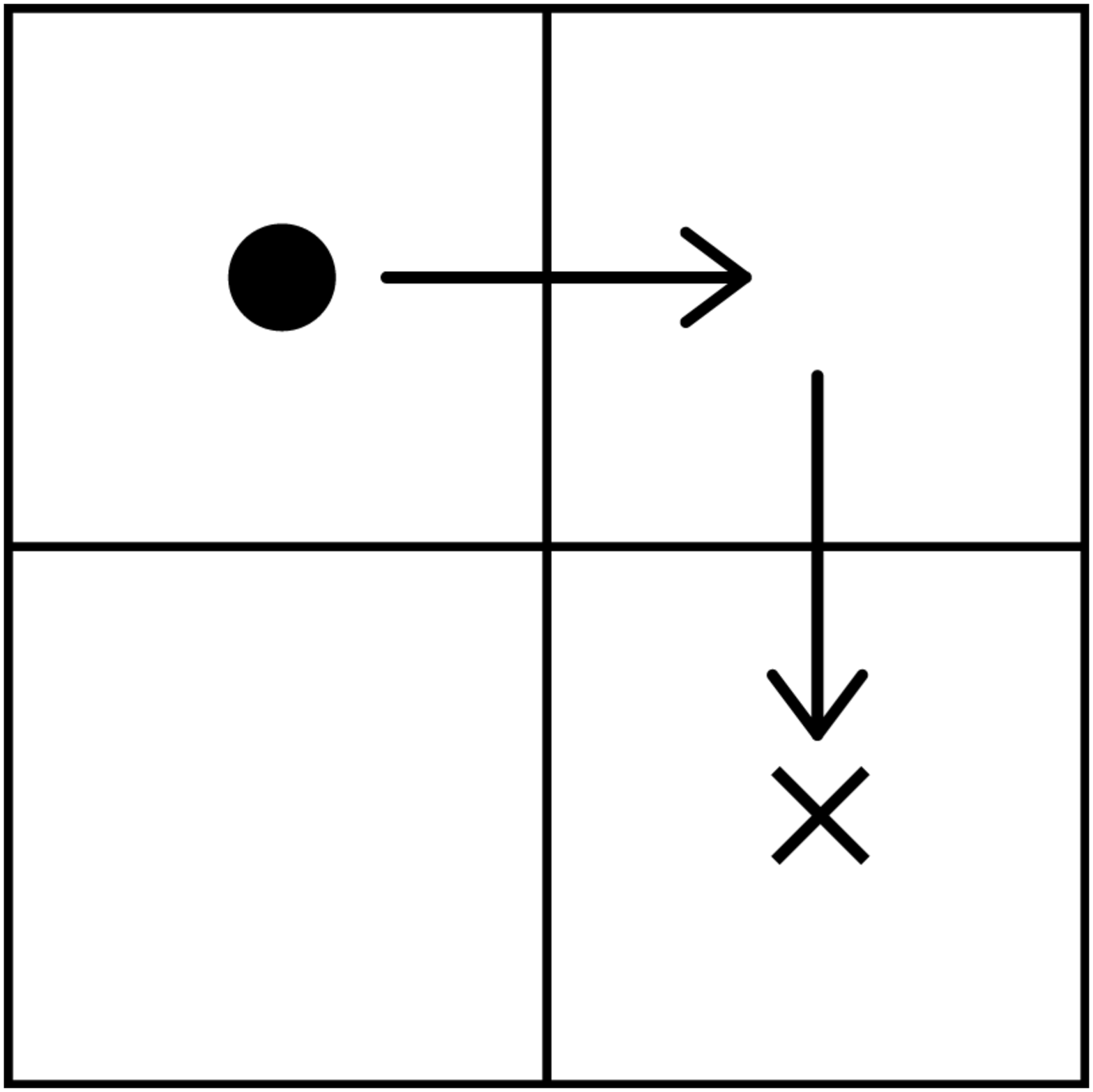} & \hspace{0.1cm} &
     \includegraphics[width=0.15\linewidth]{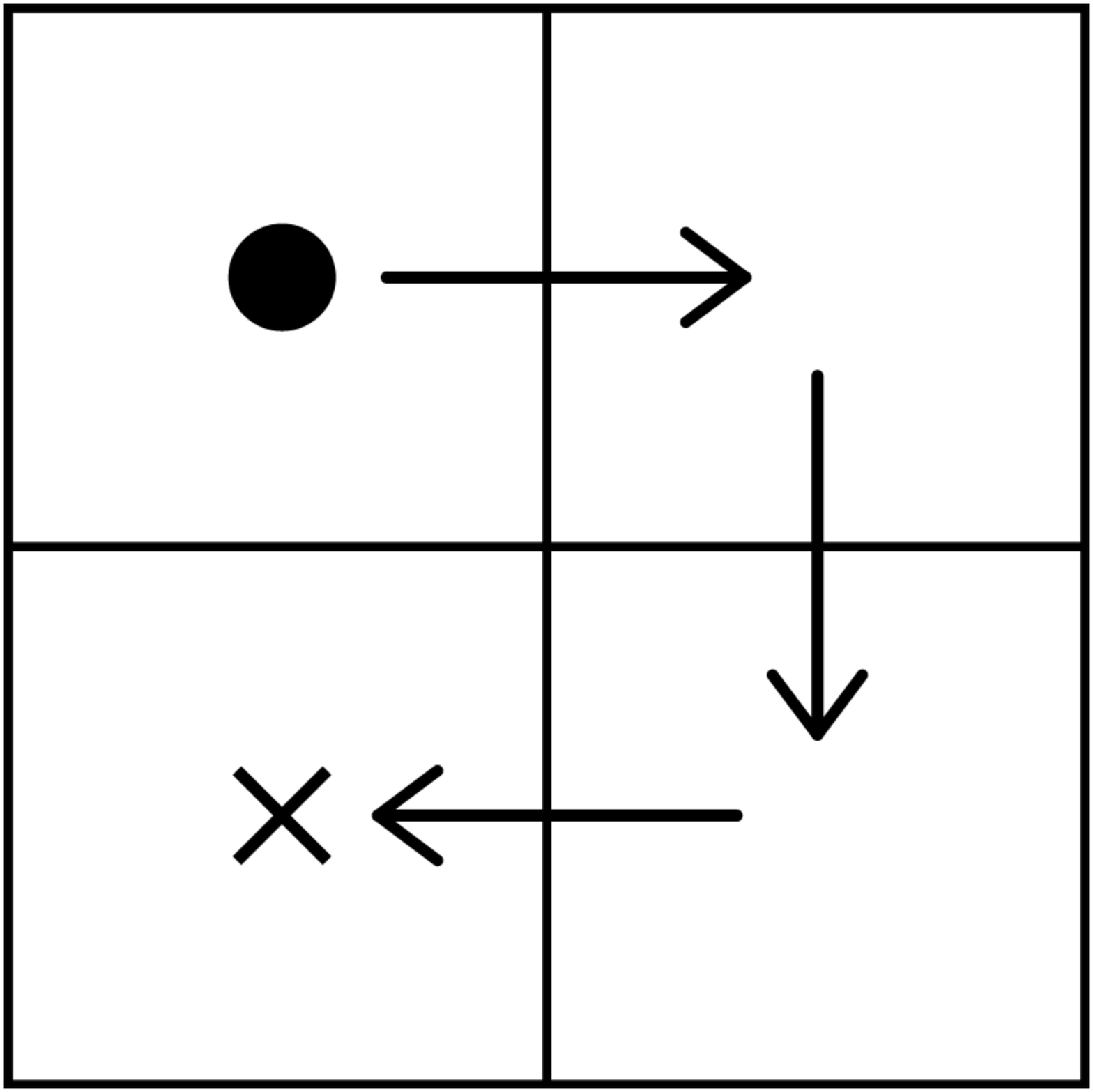} & \hspace{0.1cm} &
     \includegraphics[width=0.15\linewidth]{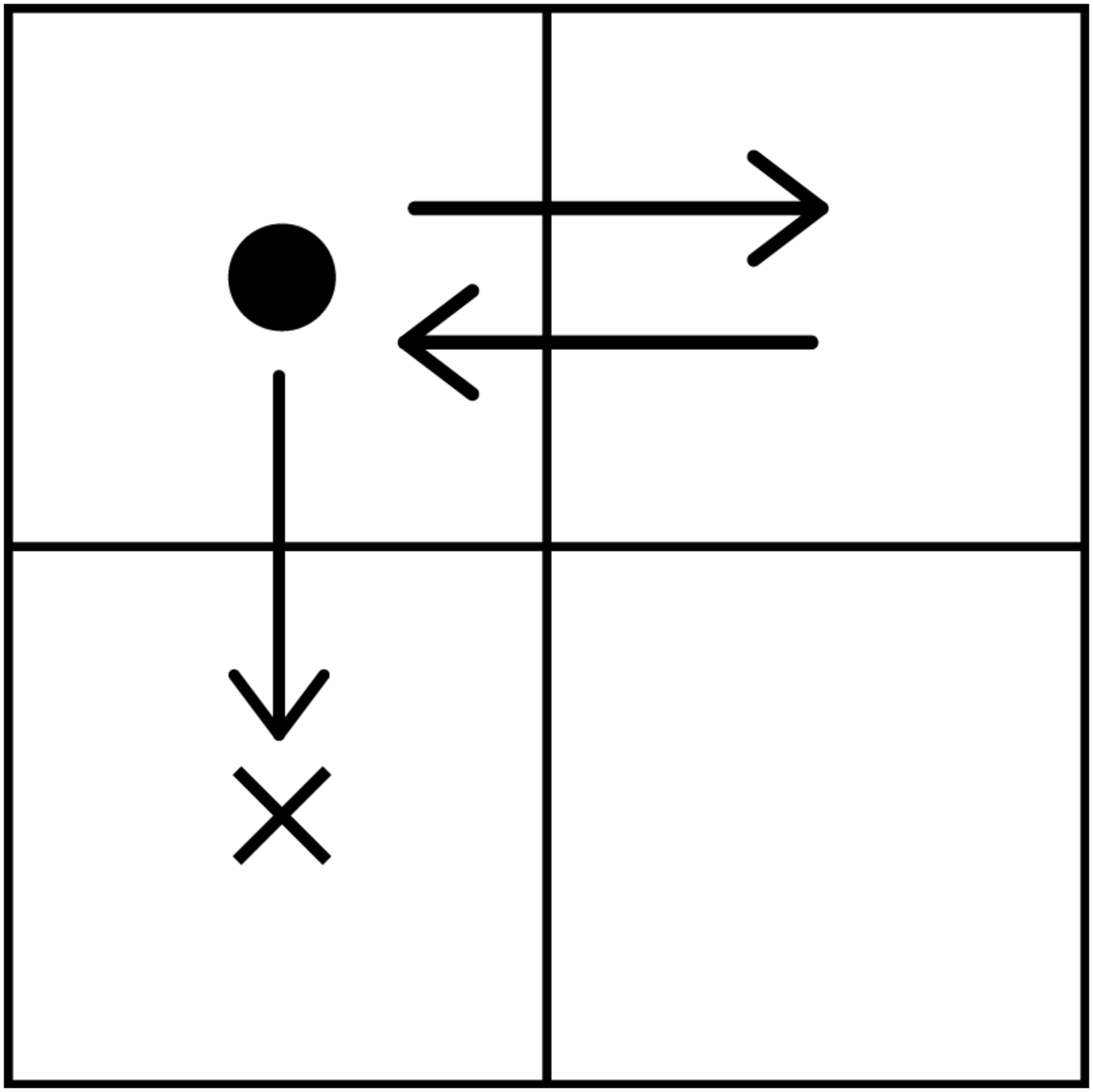} & \hspace{0.1cm} &
     \includegraphics[width=0.15\linewidth]{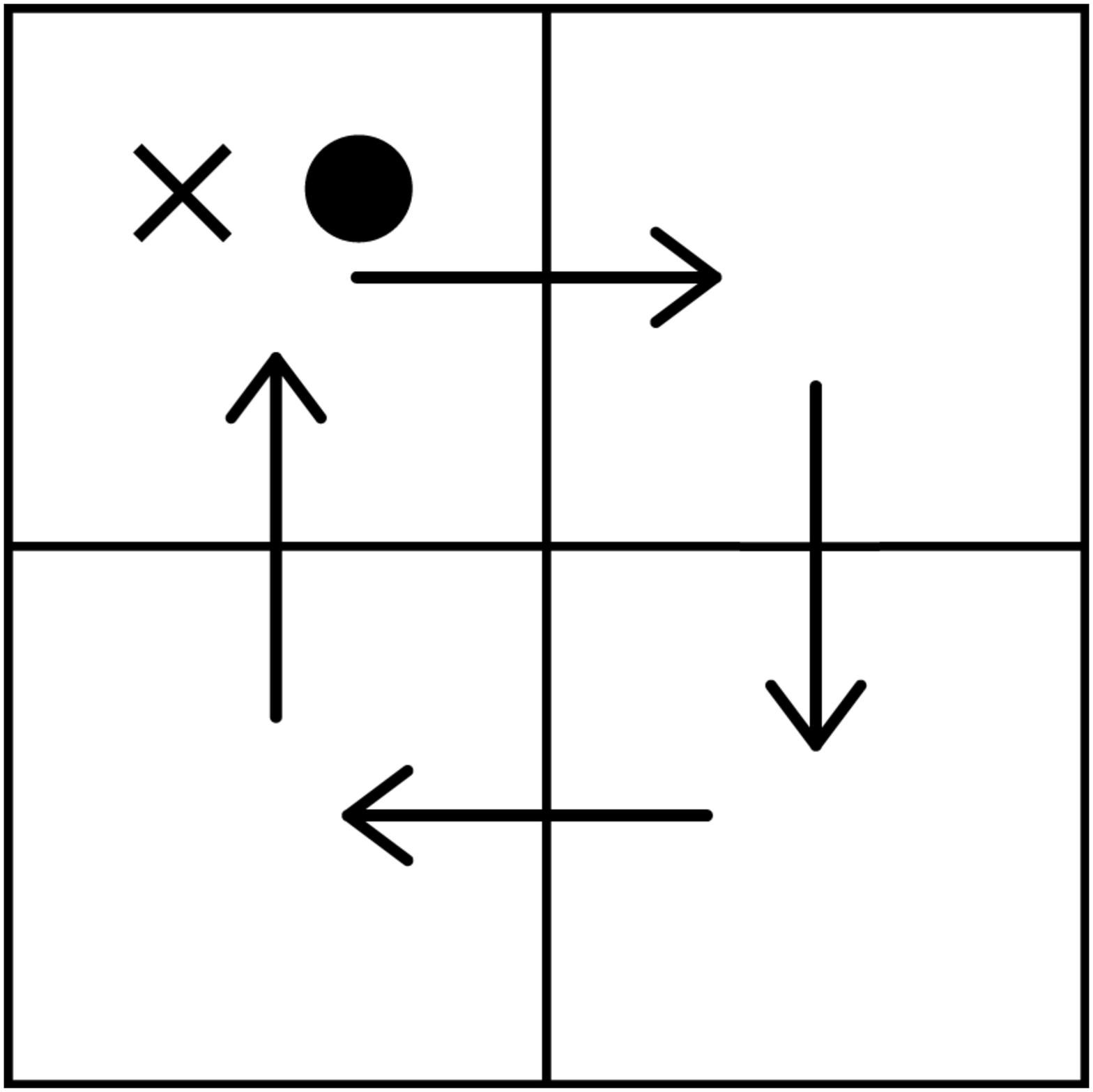} \\
      (a) & & (b) & & (c) & & (d)
    \end{tabular}
    \caption{Four examples for paths with one source ($\bullet$) and one drain ($\times$).}
    \label{fig:topology_ms}
  \end{center}
\end{figure}

The admissible digraphs can easily be completed to form a balanced connected digraph, by adding an 
extra {\it bath site}, from which a directed line points to the drain symbol and a directed line pointing from the source symbol. This is not necessarily a nearest neighbour line and can even be a self-loop. The sought-for digraphs are therefore Eulerian and can again be constructed by tracing a continuous tour. The tour begins at the origin with a source ($\bullet$) and proceeds successively along nearest neighbour dibonds. For the $n^{th}$ order term $\alpha^{(n)}_{2}$ we generate $n$ directed lines and close the graph with a drain ($\times$). (See for example \fref{fig:topology_ms} or ref.~\cite{teichmann_process-chain_2009}.)

In contrast to the energy calculation however, it is not compulsory that the path is closed. As before, the digraph is attached to the origin according to \eref{eq:shift.to.origin} and multiple copies of the same digraph are discarded.

The expectation value in \eref{eq:aux11} is invariant against relabeling the sites, i.e. isomorphic digraphs yield the same contribution to $\alpha_{2}^{(n)}$.
We again identify isomorphism by similar adjacency matrices.

In addition, \tref{tab:bh_ms_number_of_paths} shows the number of topologically unique diagrams as a function of the order $\nu$ for systems with $1$, $2$ and $3$ dimensions. The reason, why the number of topologically unique diagrams is the same for all dimensions up to third order is that with a maximum number of 3 bonds one can only draw paths that can be mapped to 1-dimensional ones (see (a), (b) and (c) in \fref{fig:topology_ms}). Only in fourth order or higher is it possible to draw a closed loop in two and three dimensions, which is not possible in one dimension (\fref{fig:topology_ms}(d)). The same reasoning applies to the 2- and 3-dimensional case, which have the same amount of topologically unique paths up to seventh order. Only in eighth order or higher paths in 3 dimensions exist that don't have topological equivalents in 2 dimensions.

\begin{table}
\caption{Number of topologically different digraphs for given order and dimension}
\label{tab:bh_ms_number_of_paths}
\begin{center}
\begin{tabular}{|r||r|r|r|}
	\hline
	order & 1-dim & 2-dim & 3-dim \\\hline 
	  1 &      1&          1 &             1 \\\hline 
	  2 &      2&          2 &             2 \\\hline 
	  3 &      4&          4 &             4 \\\hline 
	  4 &     8 &        10 &           10 \\\hline 
	  5 &     14&       22 &          22 \\\hline 
	  6 &   25 &        58 &           58 \\\hline 
	  7 &   45 &      140 &            140 \\\hline 
	  8 &   79 &      390 &           394 \\\hline 
\end{tabular} 
\end{center}
\end{table}

As before, we have to sum over all distinguishable 'time'-sequences of the elements in the digraphs, which are this time $n$ dibond hopping terms, one creation and one annihilation operator.

\begin{figure}
  \centering
    \includegraphics[width=0.8\linewidth]{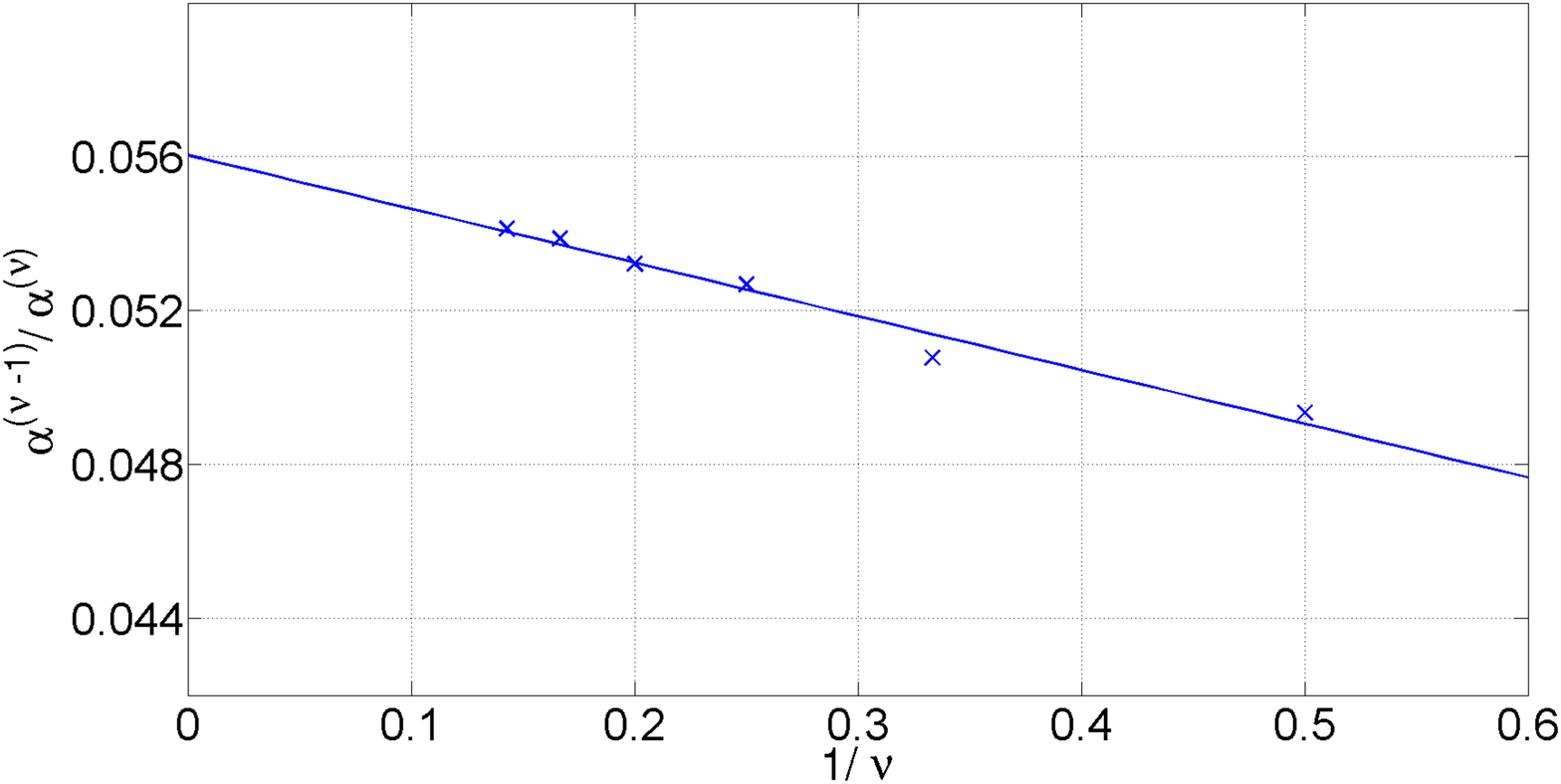}
      \caption{The ratios $|\alpha_2^{(\nu-1)}/\alpha_2^{(\nu)}|$ as function of $1/\nu$ and extrapolated to $\nu \rightarrow \infty$ using a linear fit for the first Mott lobe ($g=1$) for a 2d system with $\mu/U = 0.3$.}
  \label{fig:bh_ms_ratio_fit2}
\end{figure}

Following \eref{eq:convergence_radius}, we calculate the ratio $|\alpha_2^{(\nu-1)}/\alpha_2^{(\nu)}|$ as a function of $1/\nu$. The extrapolation $\frac{1}{\nu}\to 0$ yields the sought-for value of the 
phase boundary. In \fref{fig:bh_ms_ratio_fit2} a representative example 
is depicted. Further details and examples can be found in refs.~\cite{eckardt_process-chain_2009,teichmann_process-chain_2009}.

\subsubsection{Test results}
\label{subsec:bh_ms_results}
\begin{figure}
  \centering
    \includegraphics[width=0.8\linewidth]{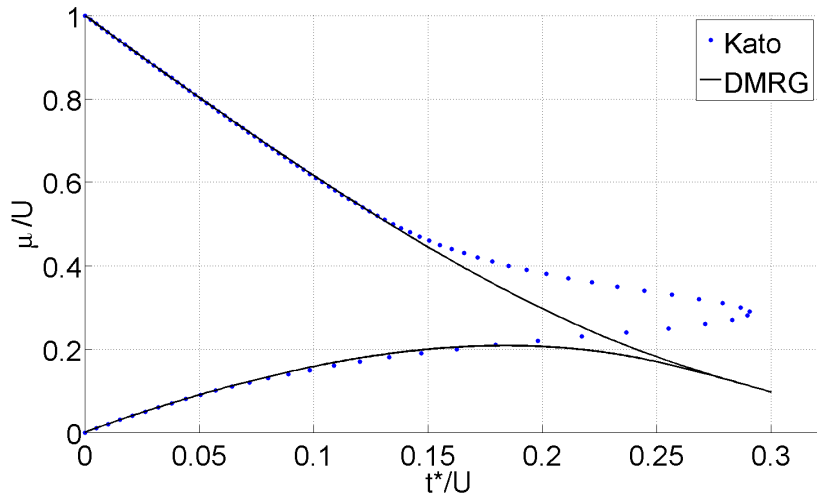}
      \caption{Comparison of the results obtained with the Kato approach using the method of effective potential and DMRG from K\"uhner et al.~\cite{kuehner_one-dimensional_2000} for a $1$-dimensional system with filling factor $g=1$.}
  \label{fig:bh_ms_comp_dmrg}
\end{figure}

The phase diagram for the BH model has already been computed with high accuracy by various techniques as well as by strong coupling approaches~\cite{teichmann_process-chain_2009,teichmann_process-chain_2009_rapid,freericks_strong-coupling_1996-1,elstner_dynamics_1999}. 
Our results agree very well with those results. 
While the 2d and 3d results agree remarkably well with results from other methods~\cite{knap_spectral_2010}, the 1d case causes problems close to the tip of the Mott lobe. In \fref{fig:bh_ms_comp_dmrg} we compare our results with data from DMRG calculations~\cite{kuehner_one-dimensional_2000}.

Teichmann et al.~\cite{teichmann_process-chain_2009} already pointed out that the strong coupling Kato-Bloch approach has some problems in detecting the phase boundary in 1d. They put the blame on the
reentrance feature of the Mott lobe: For a given chemical potential, 
the criterion based on  the radius of convergence in \eref{eq:convergence_radius} can only provide the phase boundary with the smallest $t^{*}$. We observe however, that deviations also occur for 
values of the chemical potential, where there is no reentrance feature.
The agreement with the DMRG results is very good for small values of $t$,
but close to the tip deviations start to grow.

The discrepancy in the vicinity of the tip is due to the power-law decay of the correlation functions of the Kosterlitz-Thouless (KT) transition at the tip.

The treatment of such a transition is very difficult as much depends on the extrapolation scheme. In a very recent paper, Ejima et al.~\cite{ejima_dynamic_2011} calculated the tips of the first two Mott lobes using DMRG and a suitable treatment of  the correlation functions. The results are remarkably close to those values obtained with QMC.

\subsection{Method for 1-dimensional systems}
\label{subsec:1_dimensional_systems}
As pointed out in the previous section, the power-law behaviour of the KT transistion causes problems for the method of the effective potential. We therefore incorporated another approach that allows us to determine the phase boundary for 1d systems, proposed by Freericks et al.~\cite{freericks_strong-coupling_1996-1}.

The idea is to introduce defect states, i.e.~one adds/removes one particle from the Mott ground state \mbox{$\ket{\mathbf{g}} \rightarrow \ket{\dots,g_{i-1},g_{i}\pm 1, g_{i+1},\dots}$}. With the assumption, that the compressibility approaches zero countinously at the phase boundary, the critical chemical potential $\mu^*$ can be deduced at the point where 
\begin{equation}
\Delta E^{\pm1}_{g} = \Delta E^{N \pm 1}_g - \Delta E^{N}_g = 0 \label{eq:deltaEpm1}~.
\end{equation}
This puts us in the advantageous situation that we can now use the energy algorithm described in \sref{subsec:algorithm}, which has the benefit of being immensly faster: In contrast to the method of the effective potential, we now have to consider only closed paths. The computation time can therefore be reduced by a considerable amount. Additionally, instead of having to calculate the critical $t^*$ for every $\mu$ of interest seperately, we only have to compute $\Delta E$ once for every particle (hole) band and can deduce from this all $\mu^*$ with a simple numerical operation.

The precise procedure works as follows. We start with the already well known Mott insulator ground state $\ket{\mathbf{g}}$. Then we add (remove) one particle at the site of the origin of the considered path (We shall from now on refer to this new ground state as $\ket{\mathbf{p (h)}}$). Now we let this particle (hole) hop through the lattice according to the diagram and all unique permutations of it. One has to pay attention to the fact that the system for a given diagram with $N$ involved sites is $N$-times degenerate~\cite{messiah_quantum_1999}. In other words, the state $\ket{g \pm 1, g,g}$ leads to the same unperturbed energy $\epsilon_0$ as the states $\ket{g,g \pm 1,g}$ and $\ket{g,g,g \pm 1}$.

While in the nondegenerated case, sequences containing factors of the form $S_0 \hat{h}_b S_0$ could have been omitted~(see. \eref{eq:bond_operator}), this is no longer allowed in the degenerate case.
Consequently, all Kato sequences that include a factor $\bra{\mathbf{p (h)}} \hat{h}_b \ket{\mathbf{p (h)}}$ can indeed have a nonvanishing contribution and have to be dealt with as well. Apart from that, everything works as explained in \sref{subsec:energy_corrections}.

\begin{figure}
  \centering
    \includegraphics[width=0.8\linewidth]{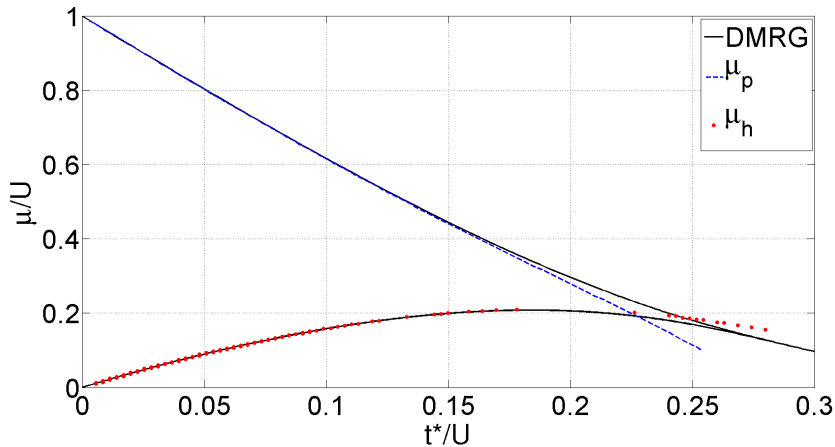}
      \caption{Comparison of the results obtained with the Kato approach and DMRG from K\"uhner et al.~\cite{kuehner_one-dimensional_2000} for a $1$-dimensional system with filling factor $g=1$.}
  \label{fig:mu_dmrg_raw}
\end{figure}

\Fref{fig:mu_dmrg_raw} shows a comparison of results calculated up to $9^{\text{th}}$ order with data from K\"uhner et al.~\cite{kuehner_one-dimensional_2000}. As one can see, the perturbation approach results match those from DMRG quite well, but deviate discernably for larger $t$, especially in the particle band case (i.e. the upper branch in the phase diagram).

\begin{figure}
  \centering
    \includegraphics[width=0.8\linewidth]{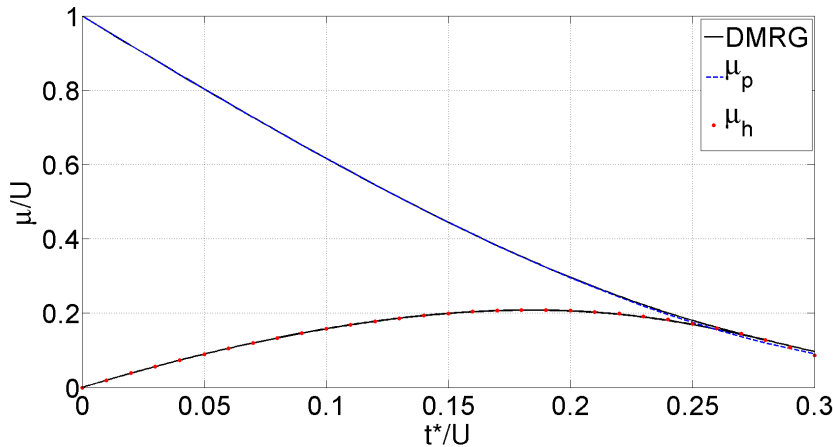}
      \caption{Comparison of our results up to $9^{\text{th}}$ order obtained with the Kato approach and DMRG from K\"uhner et al.~\cite{kuehner_one-dimensional_2000} for a $1$-dimensional system with filling factor $g=1$ after the Borel-Pad\'e transformation has been performed.}
  \label{fig:mu_dmrg_pade}
\end{figure}

This is due to the fact that the power series expansion of $\mu^*$ has an asymptotic behaviour. Calculating higher orders of the perturbation series in order to improve accuracy is therefore not feasible. Hence we are going another route and do the following:
We start with the power series for $\mu^*_p$ for the particle case (the hole case can be treated analogously), which we get from the Kato-Bloch energy calculations
\begin{eqnarray}
 \mu^*_{p}(t) &=  \sum \limits_{\nu=0}^{\nu_{max}} c_p(\nu) t^\nu ~,
\end{eqnarray}
and perform a Borel transformation (see for example ref.~\cite{hardy_divergent_2000})
\begin{eqnarray}
\mathcal{B}(z) &= \sum \limits_{\nu=0}^{\nu_{max}} c_p(\nu) \frac{z^\nu}{\nu!}~.
\end{eqnarray}
This Borel transform is now rewritten and approximated according to Pad\'e~\cite{jr_pade_1996} in order to cope with the asymptotic series.
\begin{eqnarray}
\mathcal{B}(x) &= \frac{\sum \limits_{m=0}^{m_{max}} a_m x^m}{1+\sum \limits_{n=1}^{n_{max}} b_n x^n} ~.
\end{eqnarray}
For a 1-dimensional system with filling factor $g=1$ for example we get the following series $\mathcal{B}_{p}(x)$ for the particle case and $\mathcal{B}_{h}(x)$ for the hole case (All coefficients have been rounded to two decimal figures):
\begin{eqnarray}
\mathcal{B}_{p}(x) &= \frac{1-3.11 x-3.21 x^2+1.88 x^3+1.26 x^4}{1+0.89 x-0.14 x^2-0.15 x^3+0.09 x^4} \nonumber \\
\mathcal{B}_{h}(x) &= \frac{2 x-1.75 x^2-0.02 x^3-1.27 x^4}{1+0.13 x+0.12 x^2-0.10 x^3+0.04 x^4} ~.\nonumber
\end{eqnarray}
The final series for $\tilde{\mu}^*_p(t)$ is then recovered by evaluating the integral
\begin{equation}
 \tilde{\mu}^*_{p}(t) = \int \limits_0^\infty \mathcal{B}_p(xt) e^{-x} dx ~.
\end{equation}
This revised expression for $\tilde{\mu}^*_{p}$ is very good natured and our results now agree excellently with those from the literature~\cite{kuehner_one-dimensional_2000}, as can be seen in \fref{fig:mu_dmrg_pade}.

\section{Local Disorder}
\label{subsec:disorder}
In a previous work, Freericks et al.~\cite{freericks_strong-coupling_1996-1} used Schr\"odinger-Reighleigh perturbation expansions up to third order to study a disordered BH system and Krutitsky et al.~\cite{krutitsky_ultracold_2008} further investigated such systems with high order strong coupling expansions. We will demonstrate how local disorder, described by the last term in \eref{eq:operators2} (where $\varepsilon_{i}$ is a random variable), can be dealt with in the framework of the cluster Kato-Bloch formalism. 

The effectiveness of the Kato-Bloch approach relies on reducing all paths to just topologically unique ones, which is only possible for a translational invariant system. By introducing a random disorder the invariance for a single path is broken, therefore the central quantity now has to be the grand canonical potential at zero temperature, or rather the lowest energy in Fock space for a given chemical potential, averaged over disorder realizations. By averaging over a large number of disorder realizations, the translational invariance is reobtained.
\begin{equation}
\mean{E} := \int \; E( \varepsilon) \;p(\varepsilon)\; d^{N} \varepsilon~,
\end{equation}
where $E(\varepsilon)$ is the lowest grand canonical energy for a given 
disorder realization vector $\varepsilon$ and $p(\varepsilon)$ is the 
joint probability density function (pdf) for the random disorder  variables. The latter are assumed to be 
uncorrelated and the joint pdf is the product of independent and identically distributed (iid) random variables
\begin{equation*}
p(\varepsilon) = \prod_{i=1}^{N} p(\varepsilon_{i})\;.
\end{equation*}
The $n^{\text{th}}$ order contribution to the mean energy is according to 
\eref{eq:energy_corr1}
\begin{equation}
          \mean{\Delta E^{(n)}_g} = 
          \sum_{\{k_{\nu}\}^{(n-1)}}
     \mean{
\bra{\mathbf{g}^{\varepsilon}} H_{1}  S^{\varepsilon}_{k_{n-1}} \dots  S^{\varepsilon}_{k_{1}} H_{1} \ket{\mathbf{g}^{\varepsilon}}
}~,
\end{equation}
where the ground state occupation $\mathbf{g}^{\varepsilon}$ depends on the disorder realization and will be site dependent. The same holds for the energies in the atomic limit $E^{\varepsilon}(n)$, entering the operators $S_{k}^{\varepsilon}$. The summation over all dibond sequences in \eref{eq:aux1} is again mapped into a sum over all connected and balanced labelled digraphs. Similar to  \eref{eq:graph.weight}, the weight for one digraph $\GG$ is given by
\begin{eqnarray}\label{eq:weight.graph.mean}
w(\GG) = \\
\mean{
\bra{g^{\varepsilon_{1}}_{1},\ldots,g^{\varepsilon_{m}}_{m}} \hat h_{b^{(n)}}  S^{\varepsilon}_{k_{n-1}} \dots \hat h_{b^{(2)}} S^{\varepsilon}_{k_{1}} \hat h_{b^{(1)}} \ket{g_{1}^{\varepsilon_{1}},\ldots,g_{m}^{\varepsilon_{m}}}\; } ~, \nonumber
\end{eqnarray}
where only the disorder energies of the site reached by the graph are required. The mean weight is again independent of the labeling and hence isomorphic digraphs yield the same contribution. So the algorithm is almost the same as before in the homogeneous system. We generate all topologically different, connected and balanced digraphs along with their multiplicities. For each digraph all distinct permutations of the dibonds are generated and the summation over possible Kato-Bloch sequences is performed.

For every specific dibond sequence and every set of Kato-Bloch indices the averages over the disorder realizations on the $m$ involved sites has now to be carried out.
The same modification holds for the computation of the susceptibility $\alpha_{2}^{n}$.

In order to compare with ref.~\cite{knap_excitations_2010}, we have chosen a binary disorder with 
$\varepsilon_{i}=\pm \varepsilon$, with disorder strength $\varepsilon$.

\subsection{Results}
\begin{figure}
  \centering
    \includegraphics[width=0.8\linewidth]{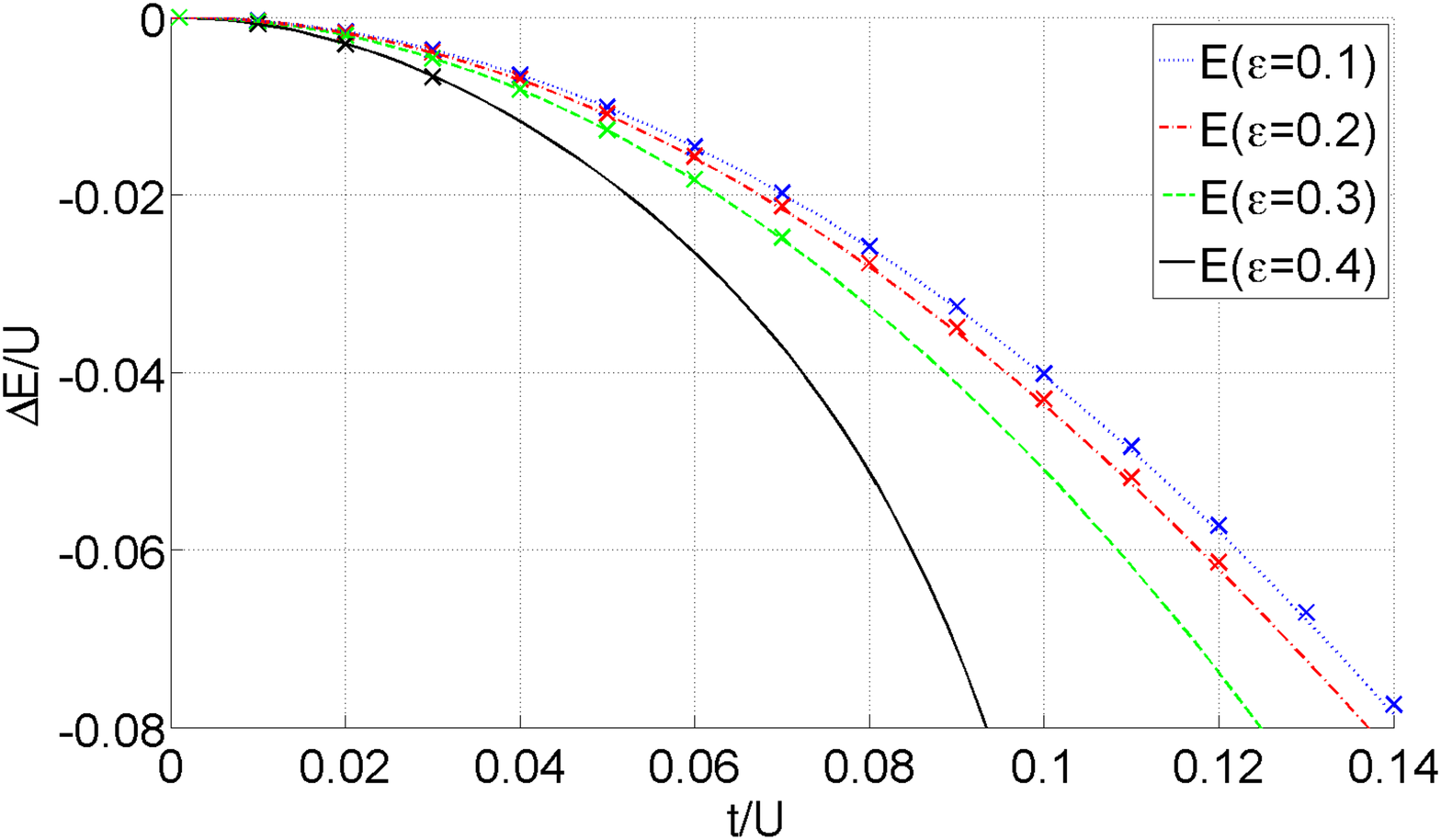}
      \caption{Energies of a $1$-dimensional system with disorder parameters $\varepsilon = 0.1$ (blue dotted line), $\varepsilon = 0.2$ (red dashed-dotted line), $\varepsilon = 0.3$ (green dashed line) and $\varepsilon = 0.4$ (black solid line) compared to results obtained with VCA~\cite{knap_excitations_2010}, which are marked with $\times$. The chemical potential is fixed at $\mu=0.5$.}
  \label{fig:bh_energy_disorder_compknap}
\end{figure}

\Fref{fig:bh_energy_disorder_compknap} shows the energy for a $1$-dimensional system as function of the hopping parameter for different disorder parameters $\varepsilon$ as well as a comparison with results obtained with VCA~\cite{knap_excitations_2010}. Energy corrections were included up to $8^{\text{th}}$ order and we averaged over 500 configurations. The solid lines are our results calculated with the Kato-Bloch algorithm and the $\times$ mark the results from VCA.

As before, our results are in excellent agreement with those computed with VCA, which can be seen especially well for the $\varepsilon=0.1$ and $\varepsilon=0.2$ cases. Unfortunately we don't have data from VCA for $\varepsilon=0.3$ and $\varepsilon=0.4$ at higher hopping parameters, so the possibilities to compare these two methods for these parameters are rather little. The data at low hopping strength and the general information gathered up to this point however suggest a good agreement of Kato-Bloch and VCA nevertheless.

\begin{figure}
  \centering
    \includegraphics[width=0.8\linewidth]{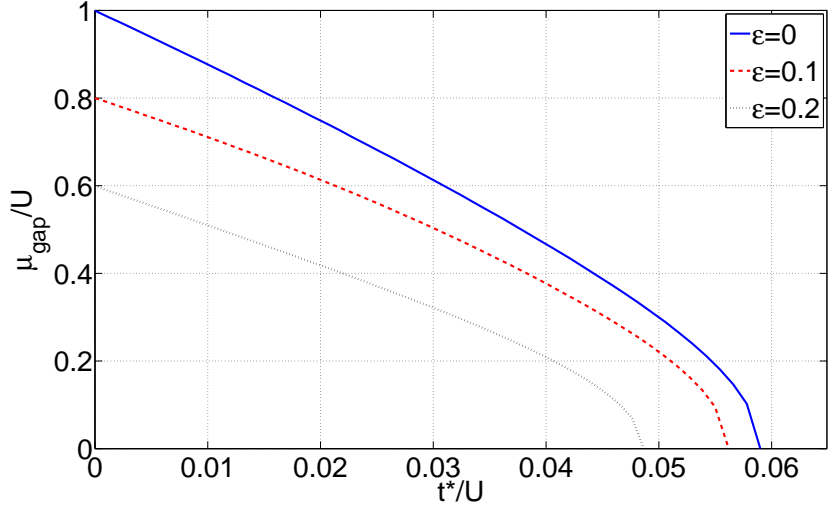}
\caption{The Mott insulator gap for a $2$-dimensional system as function of the hopping strength  for $\varepsilon=0.0$,  $\varepsilon=0.1$ and $\varepsilon=0.2$.}
  \label{fig:bh_ms_d2_mu_gap}
\end{figure}

\begin{figure}
  \centering
    \includegraphics[width=0.45\linewidth]{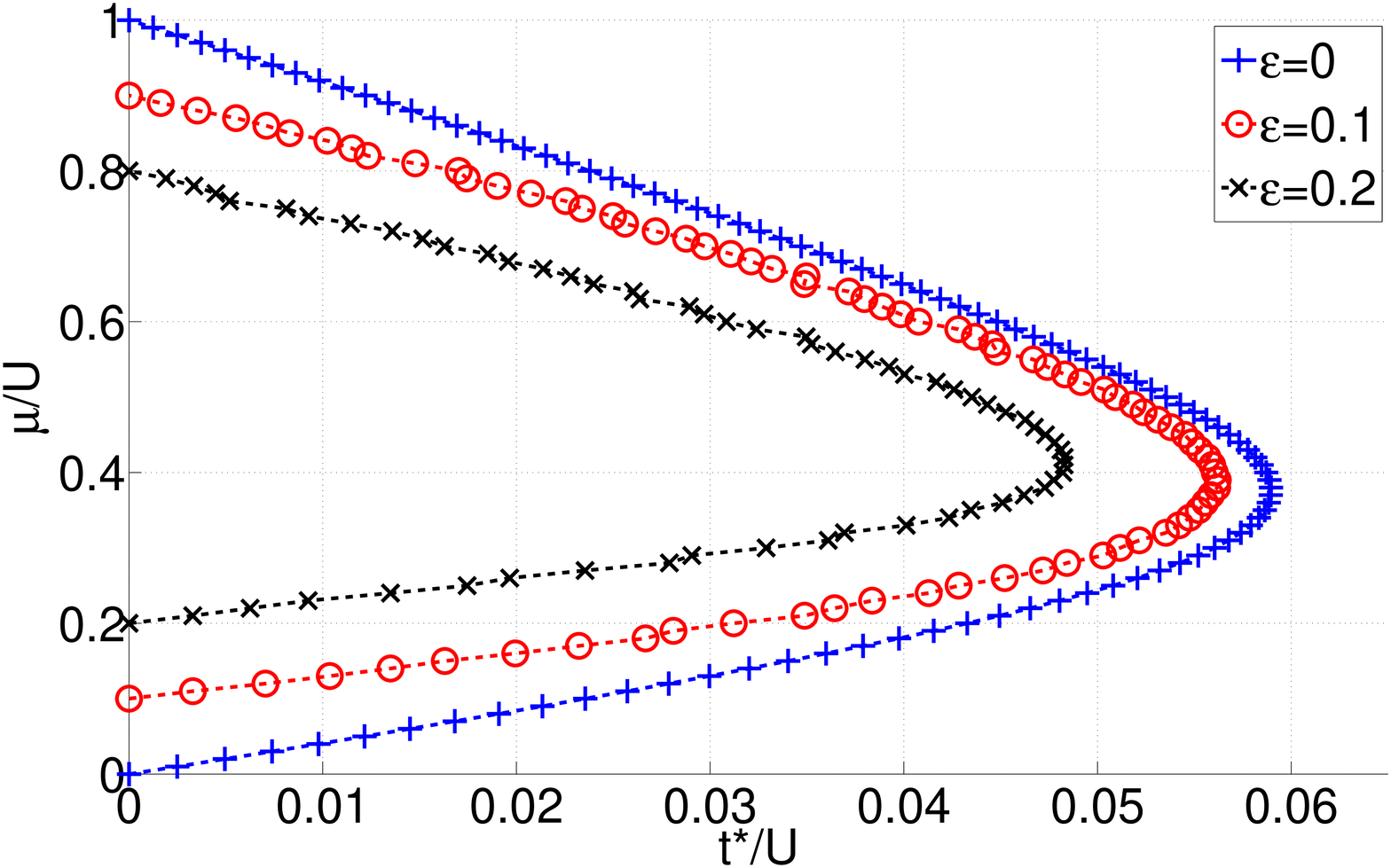}
    \includegraphics[width=0.45\linewidth]{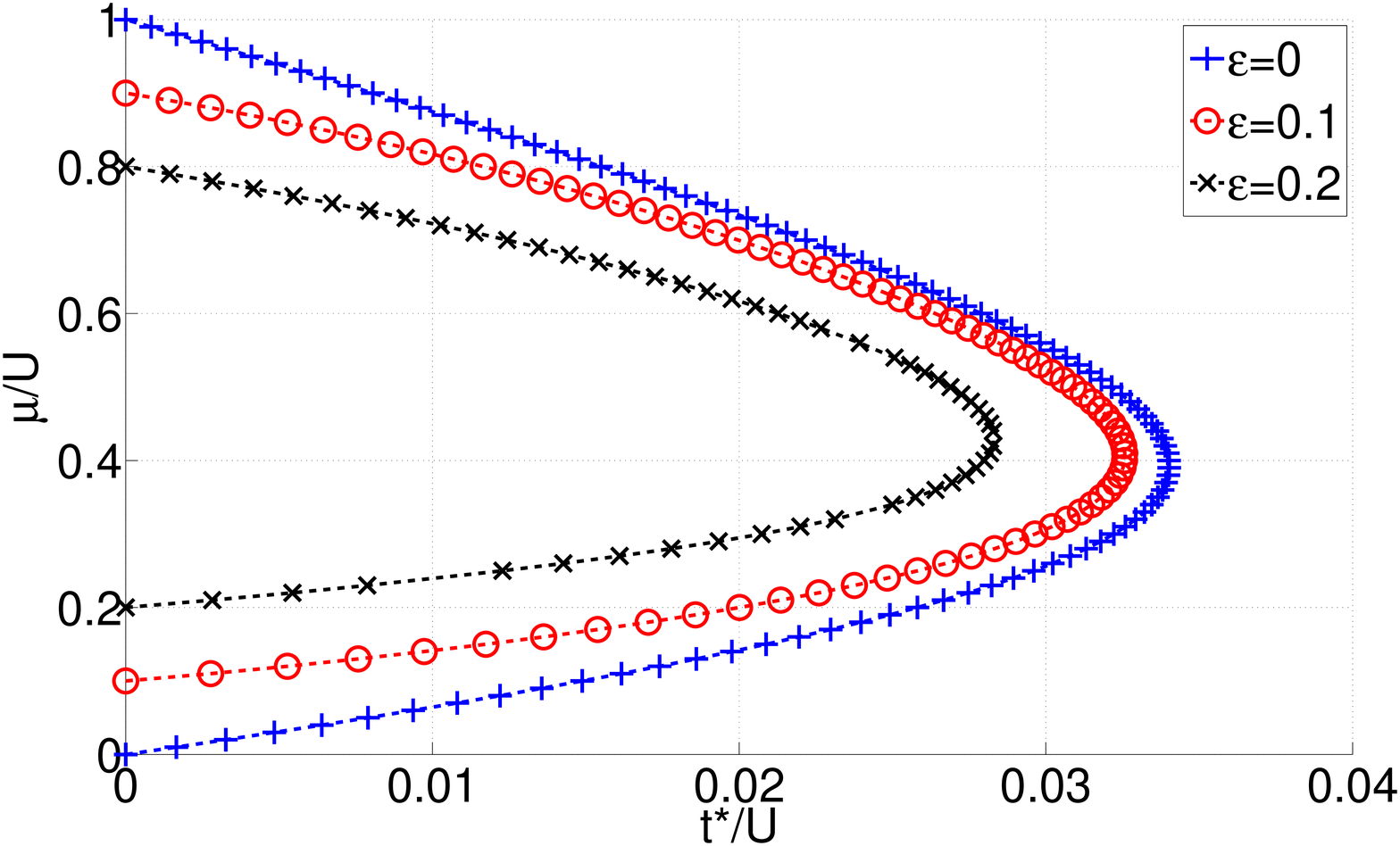}    
      \caption{Mott insulator for $g=1$ for a $2$-dimensional system (left figure)
and a $3$-dimensional system (right figure) as function of the hopping strength  for $\varepsilon=0.0$,  $\varepsilon=0.1$ and $\varepsilon=0.2$.}
  \label{fig:bh_ms_d2_phase}
\end{figure}

In \fref{fig:bh_ms_d2_mu_gap} we plotted the width of the Mott insulator regions as function of the critical hopping strength $t^*$ as a way for comparison. The blue solid line stands for the ordered system, the red one for the system with $\varepsilon=0.1$ and the black one for $\varepsilon=0.2$.
In \fref{fig:bh_ms_d2_phase} the dependence 
on the disorder parameter $\varepsilon$ is depicted for 
the $g=1$ Mott lobe. We observe that the Mott lobe shrinks with increase disorder. A comparison of the $2$-d and $3$-d reveals that the impact of the disorder is almost independent of the physical dimension.
It is important to note however, that within the present approach it is not possible to tell whether the neighbouring phase is a superfluid or a Bose glass phase.

%

\section{The Jaynes-Cummings lattice model}
\label{sec:jaynes_cummings_model}

Due to experimental progress in controlling quantum optical and atomic
systems, new  ideas have been proposed for new realizations of strongly-correlated many body systems, such as ultracold gases of atoms trapped in optical lattices
\cite{jaksch_cold_1998, greiner_quantum_2002, bloch_many-body_2008} or
light-matter systems.\cite{greentree_quantum_2006,
  hartmann_strongly_2006, hartmann_quantum_2008} 
The latter contains 
photons, that interact with atoms or atomic-like
structures. 
The interaction is strong if the photons are confined within optical
cavities.\cite{vuckovic_photonic_2003,altug_photonic_2005,altug_two-dimensional_2004} 
The coupling between photons and atoms leads to an effective 
repulsion between photons, which in turn results physical properties like in
the Bose-Hubbard model,
such as the quantum phase transition from a Mott phase, where particles are
localized on the lattice sites, to a superfluid phase, where particles
are delocalized on the whole lattice. \cite{greentree_quantum_2006}
Yet the physics of the light-matter models is far richer because two
distinct particles, namely photons and atomic-like excitations, are present.

The following short explanation follows strongly the approach of ref.~\cite{haroche_exploring_2006} and all quantities are expressed in units of $\hbar$.

The hamiltonian of a JC system consists of three parts, an atomic part $\hat{H}_a=\epsilon  \ket{ \uparrow} \bra{\uparrow}$ which assigns the energy $\epsilon$ to the excited atom states, a cavity part $\hat{H}_c= \omega_c \hat{a}^\dagger \hat{a}$ counting the bosons in the cavity and appointing them the energy $\omega_c$ and a part that describes the coupling between atom and cavity $\hat{H}_{ac}$, which can be written like

\begin{equation}
	\hat{H}_{ac} =  \frac{ \Omega}{2} (\sigma_{+}\hat{a} + \sigma_{-}\hat{a}^\dagger)~,
\end{equation}
with the Rabi frequency $\Omega$, which is a measure for the strength of the coupling of the two systems. Here, $\sigma_\pm$ corresponds to the atomic raising and lowering operators, while $\hat{a}$ and $\hat{a}^\dagger$ are the usual photonic annihilation and creation operators.

The eigenstates of $\hat{H}_a$ are $\ket{\downarrow}$, $\ket{\uparrow}$, the first denoting the ground state, the second the excited state. On the other hand, the cavity hamiltonian $\hat{H}_c$ has the eigenstates $\ket{n}$, the already known Fock-states from \sref{sec:model}. Therefore, the eigenstates of the uncoupled system will be the tensor products $\ket{n,\downarrow}$ and $\ket{n,\uparrow}$.

If the detuning of the system $\Delta = \omega_c - \epsilon$ is zero or very small compared to $\omega_c$, the states with the same particle number, i.e. $\ket{n,\downarrow}$ and $\ket{n-1,\uparrow}$ are degenerate or nearly degenerate respectively. The complete energy of a system with $n$ particles is therefore saved in a state with $n$ photons and no atomic excitation $\ket{n,\downarrow}$ and a state with $n-1$ photons and one atomic excitation $\ket{n-1,\uparrow}$. (The exception being a system with no particles, which of course can only be described by $\ket{0,\downarrow}$.) The coupling hamiltonian $\hat{H}_{ac}$ only translates between those states with the same particle number.

In this case, the energy eigenvalue can be written like
\begin{eqnarray}
	&E_{n,\pm} = n \omega_c - \frac{\Delta}{2} \pm q(n) \label{eq:jc_energy_block}~, \\
    &\text{with} \hspace{0.2cm} q(n) = \sqrt{\left(\frac{\Delta}{2}\right)^2 + n \left(\frac{\Omega}{2} \right)^2}~. \nonumber
\end{eqnarray}
The $\pm$-sign refers to the sign of $\ket{n,-}$ and $\ket{n,+}$, introduced in \eref{eq:jc_ground_state_block}~and~\eref{eq:jc_excited_state_block}.
If there are no particles in the system, there is just one state $\ket{0,\downarrow}$ possible. The energy eigenvalue in this case is $E_{n=0}=0$.

The eigenstates corresponding to the energies of \eref{eq:jc_energy_block} are
\begin{eqnarray}
	\ket{n,-} &= \cos{\Theta_n} \ket{n-1,\uparrow} - \sin{\Theta_n} \ket{n,\downarrow}  \label{eq:jc_ground_state_block}\\
	\ket{n,+} &= \sin{\Theta_n} \ket{n-1,\uparrow} + \cos{\Theta_n} \ket{n,\downarrow} \label{eq:jc_excited_state_block}~,
\end{eqnarray}
where we used the short notations
\begin{eqnarray}
	&\sin{\Theta_n} = \sqrt{ \frac{q(n)-\frac{\Delta}{2}}{2 q(n)}} \qquad \text{and} \\ \nonumber
	&\cos{\Theta_n} = \sqrt{ \frac{q(n)+\frac{\Delta}{2}}{2 q(n)}}~.
\end{eqnarray}
Additionally to what was said in the previous paragraphs, an atom-cavity site may also have an additional energy which is dependend on the total particle number, due to a chemical potential $\mu$. The final JC hamiltonian for an atom-cavity system looks as follows:
\begin{eqnarray}
    \hat{H}_{JC}=&\omega_c \hat{a}^\dagger \hat{a} + \epsilon \ket{ \uparrow} \bra{\uparrow  }+ \\
    &+\frac{\Omega}{2}(\hat{a}  \ket{\uparrow} \bra{\downarrow} + \hat{a}^\dagger  \ket{\downarrow} \bra{\uparrow}) - \mu(\hat{a}^\dagger \hat{a} + \ket{ \uparrow} \bra{\uparrow})~. \nonumber \label{eq:jaynes_cummings_hamiltonian_site}
\end{eqnarray}
Now we are going to construct a regular lattice by arranging a large array of such sites and allow the bosons to tunnel to 
neighbouring sites with tunneling strength $t$. The JC hamiltonian of a single site $\hat{H}_{JC,i}$ is given in \eref{eq:jaynes_cummings_hamiltonian_site} and the hopping hamiltonian is the same as for the BH model (see \sref{sec:model})
\begin{equation}
	\hat{H}_{JCL} = \sum\limits_i \hat{H}_{JC,i} + \hat{H}_{hop}~.
\end{equation}

\subsection{Changes in the algorithms}
\label{subsec:changes_algorithms_jc}
The JC site hamiltonian $\hat{H}_{JC}$ does not alter the particle number of the eigenstates, but the hopping hamiltonian $\hat{H}_{hop}$ does. It is therefore clear that the hopping term is again the perturbative part.

The fact, that there are two eigenstates of the JC model now belonging to the same particle number $n$ involves a disadvantage. As explained before, the term $S_k$ of our Kato formula projects the current state onto the according eigenstate of $\hat{H}_0$, i.e. it will either be projected onto $\ket{n,-}$ or $\ket{n,+}$. This means, that one has to add the information to which eigenstate we are projecting, i.e. $S_k \rightarrow S_{k,\sigma}$ with $\sigma$ indicating which eigenstate we should take:
\begin{equation}
      S_{k,\pm} = \cases{ 
      -\delta_{n,g} \delta_{\sigma,-} \hspace{2cm} \qquad \text{for } k=0  \hspace{2cm} \\ 
    \frac{ 1- \delta_{n,g} \delta_{\sigma,-} }{\left( E^{(0)}_{g,-} - E^{(0)}_{i,\pm} \right)^k} \hspace{1.7cm} \text{otherwise.}} \label{eq:def_S_jc}
\end{equation} 
In order to calculate the energy correctly we also have to consider not only every possible sequence $\{ k_{\nu}\}^{(n-1)} $ but also every possible permutation of the signs $\{ \sigma_{\nu}\}^{(n-1)}$.

\begin{figure}
  \centering
      \includegraphics[width=0.1\textwidth]{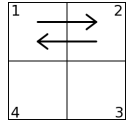}
      \caption{One possible path in second order.}
  \label{fig:paths2a_raster_numbered}
\end{figure}

Our equation for the complete energy correction of $n^{\text{th}}$ order therefore reads
\begin{equation}
\label{eq:energy_corr_jc}
	\Delta E^{(n)}_g = \sum\limits_{\{ \sigma_{\nu}\}^{(n-1)}} \Delta E^{(n)}_{g,\mathbf{\sigma_\nu}}~.
\end{equation}   
To deduce the right $\sigma_\nu$ sequences we will take a look at a few different paths. The most simple path appearing in the second order energy correction is depicted in \fref{fig:paths2a_raster_numbered}.
\begin{figure}
  \centering
      \includegraphics[width=0.22\textwidth]{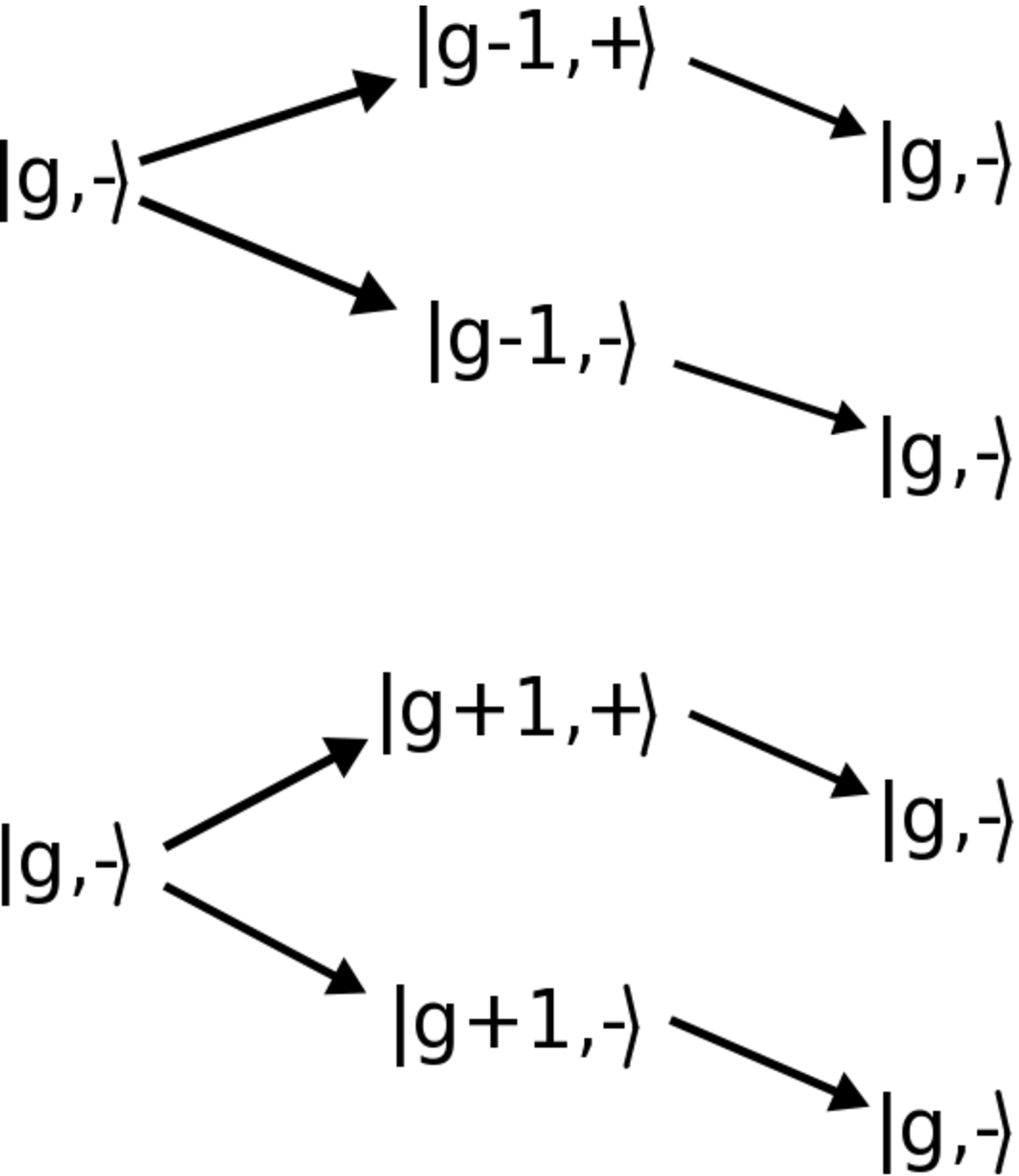}
      \caption{All possible sequences of the path in \fref{fig:paths2a_raster_numbered}.}
  \label{fig:raster_2}
\end{figure}
%
After the first hop from site $1$ to site~$2$, we do have $g-1$ particles at site~$1$. But the state of this site could have either been projected to $\ket{g-1,-}$ or $\ket{g-1,+}$. An exception would of course be if there were $g=1$ particles at each site before the hopping, in which case only the $\ket{0,\downarrow}$ state would be possible for site~$1$ after the first hopping. We will from now on omit mentioning this exception as it always results in the same state.

Site $2$ contains $g+1$ particles after the first tunneling and will be projected either onto the $\ket{g+1,-}$ or the $\ket{g+1,+}$ state. After the second hopping takes place, both sites are again populated by $g$ particles and at each site the ground state $\ket{g,-}$ has to be present in order to lead to a non-vanishing energy contribution. A graphical depiction of this can be seen in \fref{fig:raster_2}.

To calculate the energy contribution of this path correctly, we would now have to calculate the energy of each of the $4$ possible sequences and sum them up.

\begin{figure}
  \centering
      \includegraphics[width=0.1\textwidth]{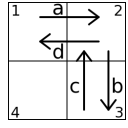}
      \caption{One possible path in fourth order with hopping sequence \{1,2-2,3-3,2-2,1\}.}
  \label{fig:path4perma_raster_numbered}
  \end{figure}

\begin{figure}
  \centering
      \includegraphics[width=0.35\textwidth]{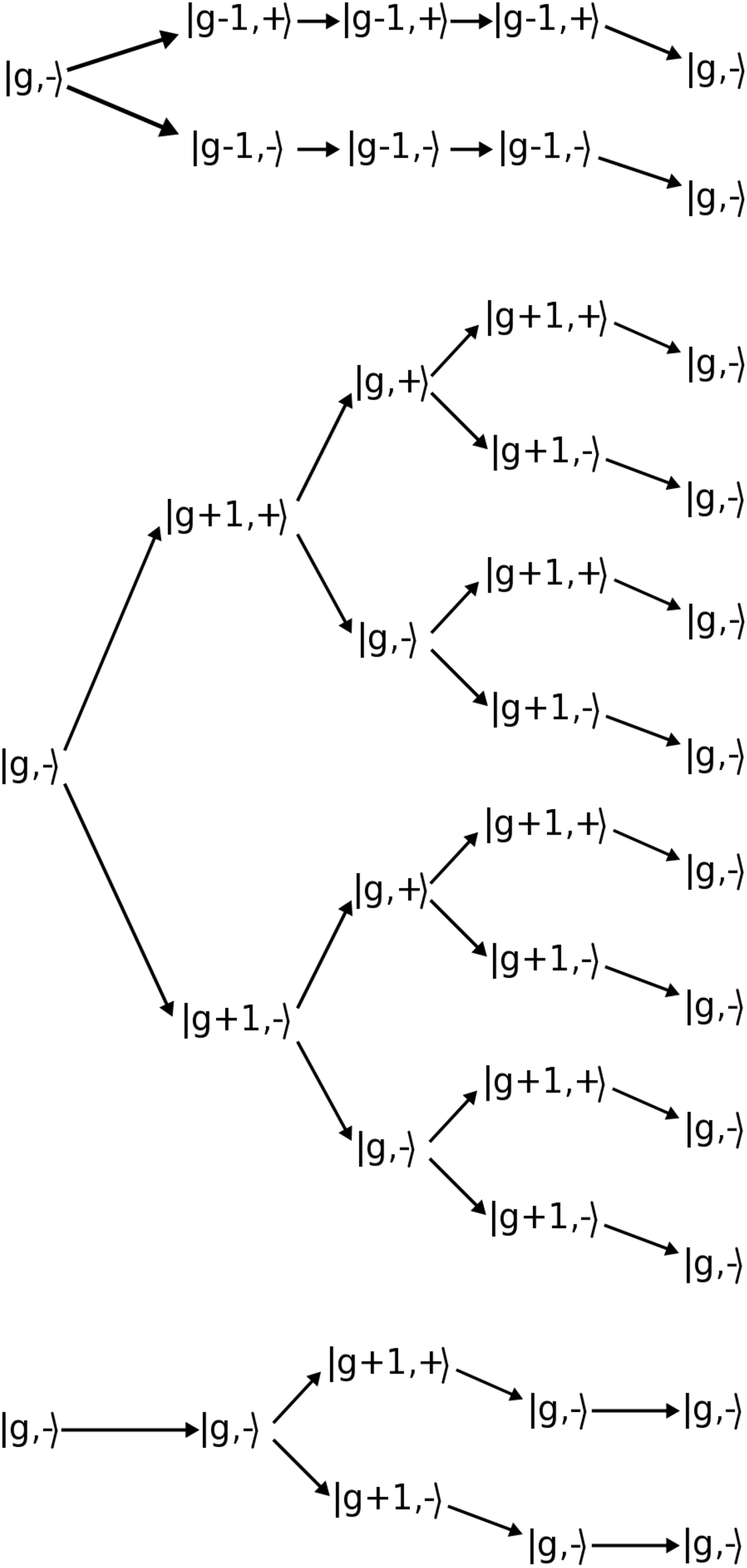}
      \caption{All possible sequences of the path in \fref{fig:path4perma_raster_numbered}}
  \label{fig:raster_4a}
\end{figure}

\begin{figure}
  \centering
      \includegraphics[width=0.1\textwidth]{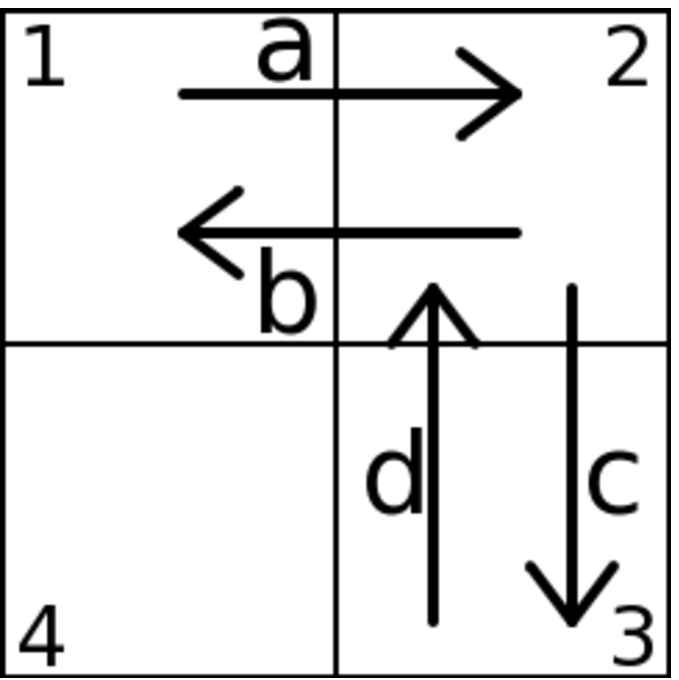}
      \caption{One possible path in fourth order with hopping sequence \{1,2-2,1-2,3-3,2\}.}
  \label{fig:path4permb_raster_numbered}
  \end{figure}

\begin{figure}
  \centering
      \includegraphics[width=0.3\textwidth]{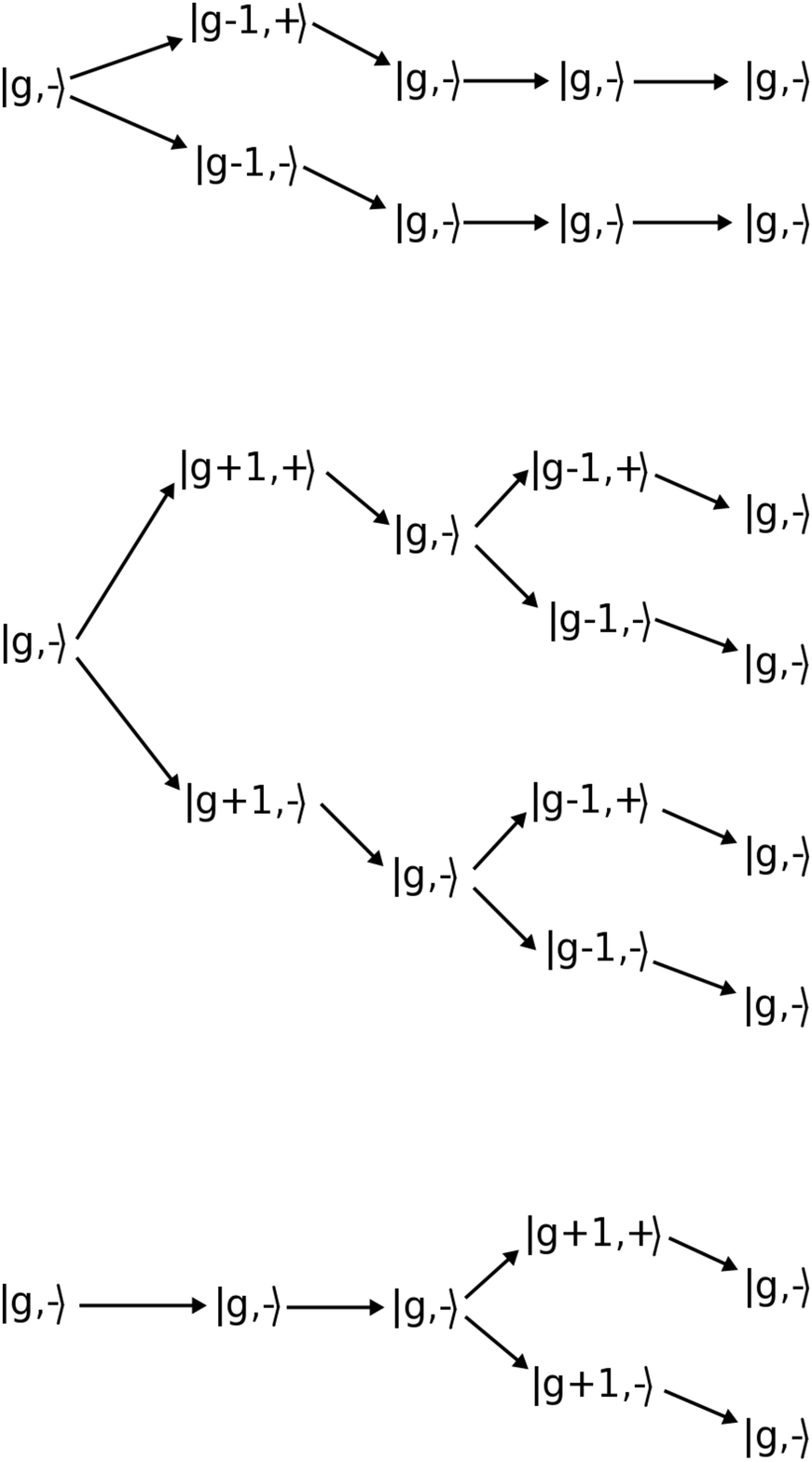}
      \caption{All possible sequences of the path in \fref{fig:path4permb_raster_numbered}}
  \label{fig:raster_4b}
\end{figure}

In order to get the right number of sequences, we have to include information about the number of nodal points, i.e. how many times the ground state is restored before the whole perturbation process is finished. An instructive example is shown in \fref{fig:path4perma_raster_numbered} and \fref{fig:path4permb_raster_numbered}. While both of these figures show the same diagram, the order of hoppings is different. In \fref{fig:path4perma_raster_numbered} the ground state is never present up until the very end resulting in the sequences of \fref{fig:raster_4a}. For the path depicted in \fref{fig:path4permb_raster_numbered} however we assume it restores the ground state after two hopping processes, leading to the sequences of \fref{fig:raster_4b}. (If the path of \fref{fig:path4permb_raster_numbered} does not result in the ground state after the first two hoppings but in the $| \mathbf{g} ,+ \rangle$ state, we would get a very similar diagram for the sequences as that of \fref{fig:raster_4a}.)
%
To compute the total numbers of sequences for a certain path, we count the number of arrows pointing to a site $N_{a,i}$. For closed paths this number has to be of course the same as the number of arrows pointing away from a site, the number of bonds of \mbox{site $i$} is therefore $N_{B,i}=2N_{a,i}$. Additionally, we need to know the number of nodal points $N_{n,i}$ for this site. The number of sequences $N_{seq,i}$ for \mbox{site $i$} is then
\begin{eqnarray}
	N_{seq,i} &= 2^{N_{B,i}-(N_{n,i}+1)}~,
\end{eqnarray}
and the total number of sequences $N_S$ is in further consequence the product of these $N_{seq,i}$
\begin{eqnarray}
	N_{S} &= \prod\limits_i N_{seq,i}~.
\end{eqnarray}
The number of bonds for a specific site can be determined very easily with the aid of adjacency matrices. The $i^{\text{th}}$ row contains the number of bonds originating from site $i$. The $i^{\text{th}}$ column on the other hand contains the information from which sites bonds point to site $i$. The adjacency matrix for the path shown in \fref{fig:path4perma_raster_numbered} is
\begin{equation}\label{eq:adjacency_bonds}
M = 
\left(\begin{array}{ccc}
0 & 1 & 0 \\
1 & 0 & 1 \\
0 & 1 & 0  
\end{array}\right)~.
\end{equation} 
The vector $v_B$ resulting when summing over all columns of the matrix stated by \eref{eq:adjacency_bonds} reads $v_B^T=(1,2,1)$. Doubling it results in $(2,4,2)$, which means that there are two bonds at site $1$, four bonds at site $2$ and again two bonds at site $3$.

With all that in mind we can reuse most of the algorithms created for the energy calculations of the BH model. The path creation and the reduction to topologically unique diagrams can be copied without any change. The algorithm for the energy calculation however has to be adopted at the point where all the permutations of a diagram are created. At this point we have to compute the adjacency matrix for every permutation to determine the number of bonds for each site. With this information we can create the allowed $\sigma$-sequences for each permutation individually, calculate the energy for every sequence and sum them up. After all $\sigma$-sequences have been worked off, we proceed to the next permutation.

\subsection{Results}
\label{subsec:jc_energy_results}
The additional summation over all $\sigma$-sequences of course slows down the energy computation compared to the calculations for the BH model, but the computation times up to order $6$ are still in the range of seconds, as \tref{tab:jc_energy_times} shows.
\begin{table}
\caption{Computation times in seconds for the JCL energy correction up to order $\nu=6$ for different filling factors $g$ and dimensions $d$. (All computations were done with Matlab on a computer with a clock frequency of 3 GHz.)}
\label{tab:jc_energy_times}
\begin{center}
\begin{tabular}{c||r|r|r||r|r|r}
           &              &   $g=1$  &                &                & $g=2$       &   \\\hline
$\nu$ & $d=1$  & $d=2$    & $d=3$    & $d=1$    & $d=2$       & $d=3$     \\\hline \hline
  $ 2$ & $0,25$ &   $0,24$ &  $0,35$  &  $ 0,23$ &    $  0,22$ &    $  0,32$ \\\hline
   $4 $& $0,37$ & $  0,46$ &  $ 0,53$ &  $ 0,58$ &   $   0,72$ &   $   0,75$ \\\hline
   $6$ & $8,30$ & $27,15$ & $27,71$ & $58,61$ & $153,06$ & $156,54$
\end{tabular}
\end{center}
\end{table}

All systems we are going to discuss have in common that we set the detuning $\Delta=0$. A more detailed look at different sets of parameters can be found in ref.~\cite{koch_superfluid_mott-insulator_2009}.

\begin{figure}
  \centering
    \includegraphics[width=0.8\linewidth]{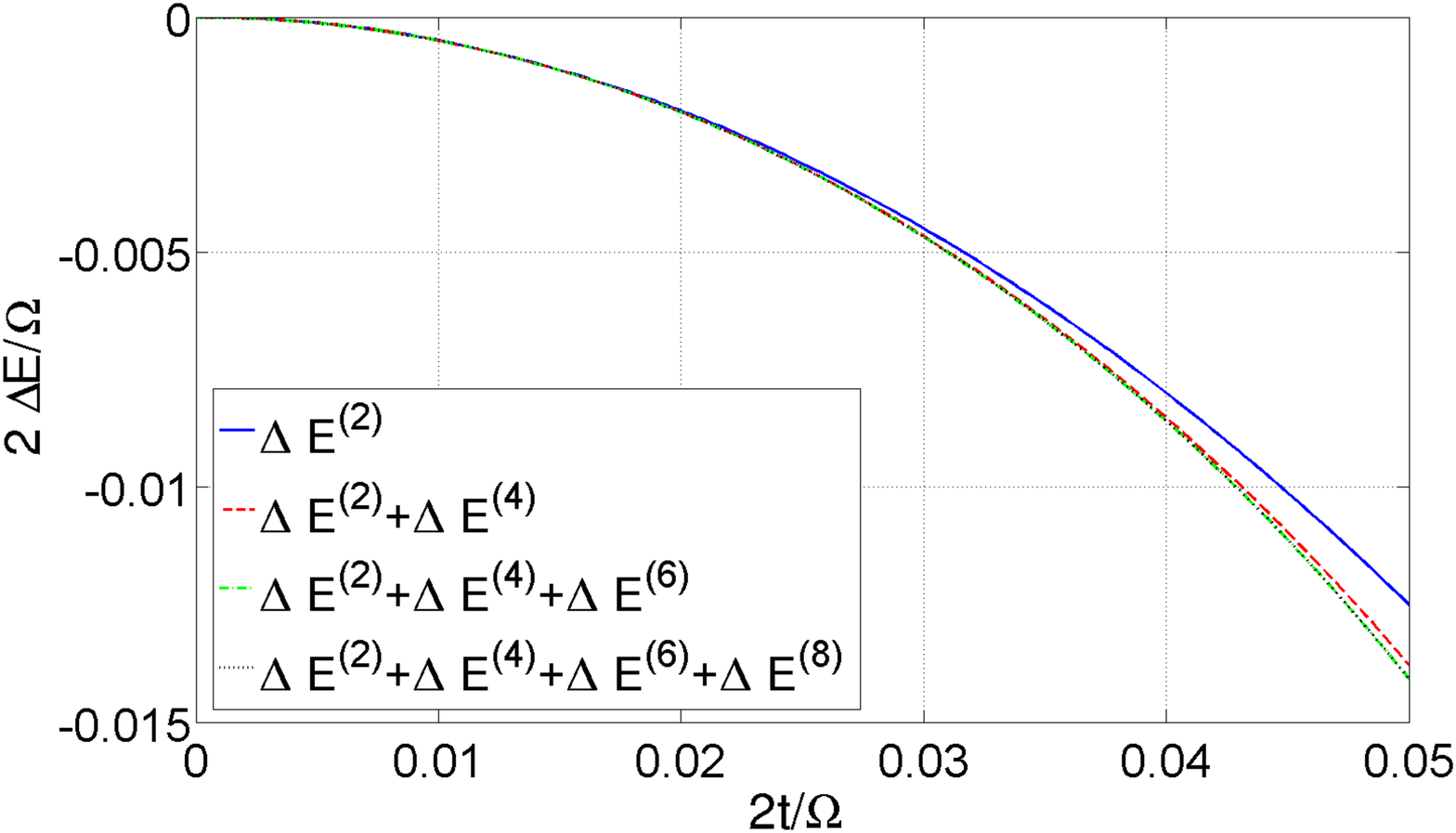}
      \caption{Comparing the fidelity of the computed energy corrections depending on the orders of correction included for a JCL system with $2$ dimensions and unity filling factor.}
  \label{fig:jc_energy_fidelity}
\end{figure}

\Fref{fig:jc_energy_fidelity} shows the computed energy correction of a $2$-dimensional system with filling factor $g=1$ for different orders of correction. The solid (blue) line contains only the second order energy correction $\Delta E^{(2)}$, the dashed (red) line results from combining the second order and fourth order corrections, the dashed-dotted (green) line shows the energy correction up to sixth order and the dotted (black) line up to the eighth order.

While the solid (blue) line differs from the others quite significantly, the dashed and dotted lines are very close to each other. In fact, the lines for the $6^{th}$ and $8^{th}$ order lie on top of each other perfectly at this scale, suggesting a very high fidelity already for the sixth order correction.

\begin{figure}
  \centering
    \includegraphics[width=0.8\linewidth]{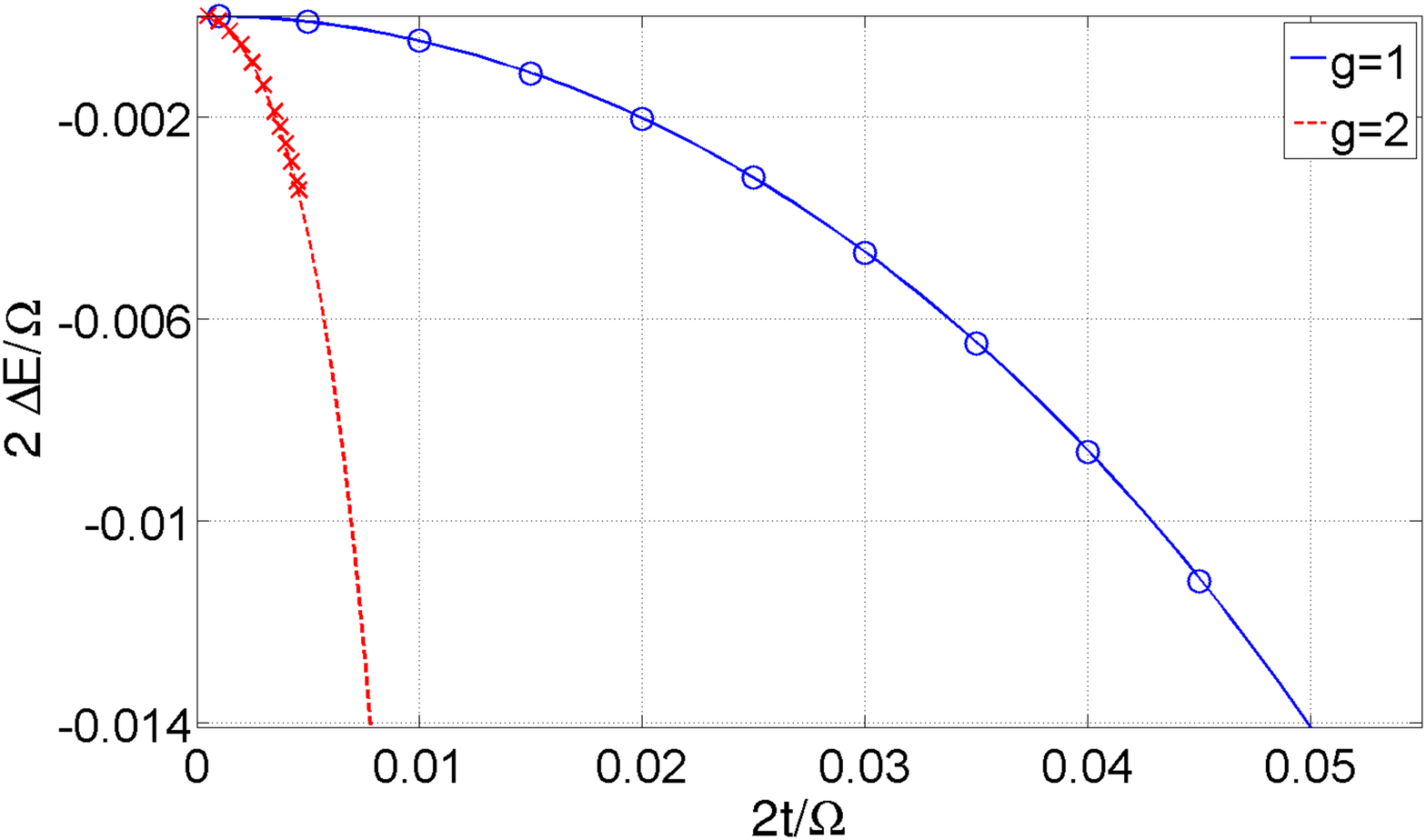}
      \caption{Comparison of calculations carried out with the Kato-Bloch algorithm (solid and dashed lines) and with VCA~\cite{knap_polaritonic_2011} (circles and $\times$) of two systems with dimension $d=2$ but different fillings factors $g$.}
  \label{fig:jc_energy_comp_knap}
\end{figure}

Additionally, \fref{fig:jc_energy_comp_knap} shows a comparison of data computed with VCA~\cite{knap_polaritonic_2011}. Two systems were simulated, with the first having $2$ dimensions and unity filling factor and the second also $2$ dimensions but filling factor $g=2$.

The solid and the dashed lines represent the results obtained with the Kato-Bloch approach and the \mbox{circles and $\times$ those} from VCA. As previously with the BH system (see \fref{fig:energy_comp_knap} in \sref{subsec:bh_results}), our results are in excellent agreement with VCA.

\begin{figure}
  \centering
    \includegraphics[width=0.8\linewidth]{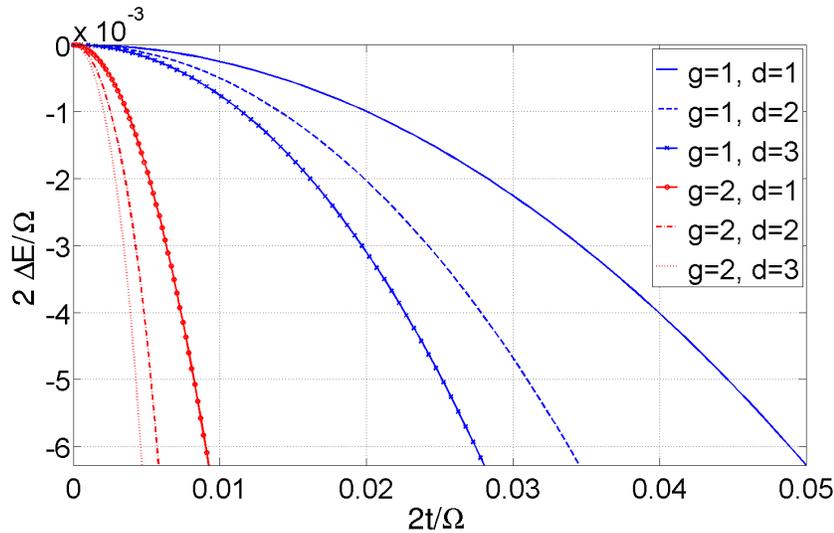}
      \caption{Comparison of the energies of systems with various filling factors~$g$ and dimensions $d$ as a function of the hopping strength $t$.}
  \label{fig:jc_energy_comp_g_d}
\end{figure}

A comparison of the energy of different systems with $1$, $2$ and $3$ dimensions and with the filling factors $g=1$ and $g=2$ is shown in \fref{fig:jc_energy_comp_g_d}.
Based on the numerical results, it appears that the energies depend stronger on the filling factor $g$ than on the physical dimension $d$.

\subsection{The Mott Insulator-Superfluid Phase Transition}\label{subsec:jc_mott_insulator_superfluid_transition}
\begin{figure}
  \centering
    \includegraphics[width=0.8\linewidth]{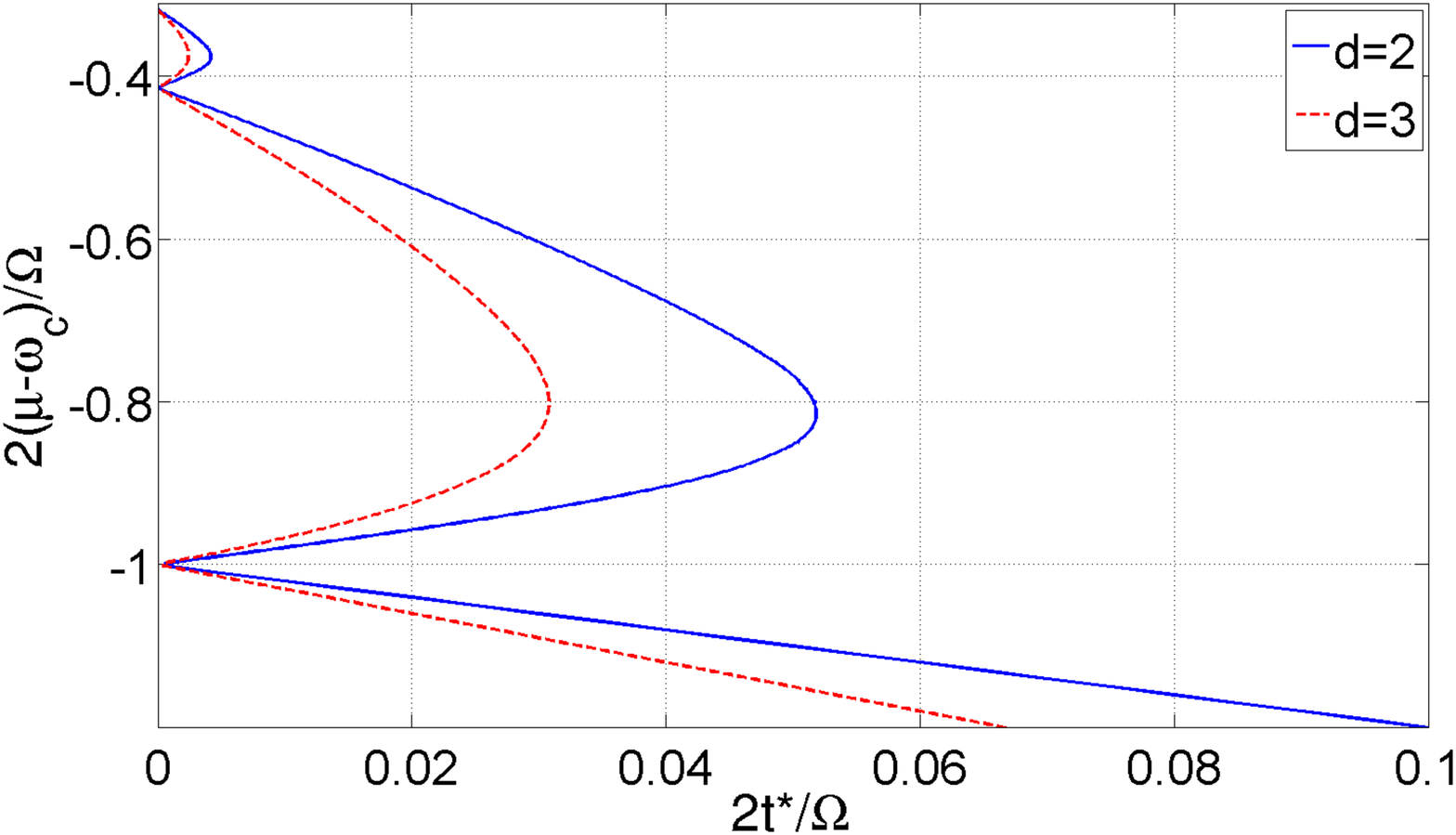}
      \caption{Boundary of the Mott phase  for $d=2$ and $d=3$. 
      The detuning is zero $\Delta=0$.  The filling factors for the three Mott lobes
      are $g = (0,1,2)$ from the bottom to top.}
  \label{fig:jc_ms_comp_d_g}
\end{figure}

The same changes in the algorithm that have been discussed in the previous chapter have also to be taken into account if one wants to determine the Mott insulator-superfluid phase boundary for the JCL model. But furthermore it is no longer necessarily true that the number of bonds pointing to a site is the same as the number of bonds pointing away from it. This little inconvenience is taken care of by just computing the sum along the rows and the sum along the columns of the adjacency matrix and adding these vectors point-wise, thereby getting the number of bonds per site.
In \Fref{fig:jc_ms_comp_d_g} the Mott insulator-superfluid phase boundary is shown for $d=2$ and $d=3$ and different filling factors. Left to the depicted lines, the system is in the Mott phase.

The phase boundary $\mu^{*}_{0}$ for $t^*=0$ is independent of the dimension of the system as it is determined by a single site. It can be calculated easily, resulting in $\mu^{*}_{0}=\sqrt{g}-\sqrt{g+1}$ (see ref.~\cite{jaynes_comparison_1963,koch_superfluid_mott-insulator_2009}). 
This 
A comparison with the result for the BH model, where $\mu^{*}_{0}=U g$, reveals that the effective Hubbard interaction $U$ in the Jaynes-Cummings model depends on the filling fractions. This explains, why the Mott lobes in the JC model decrease rapidly with filling fraction.

As was the case with the BH model, the phase boundary for 1-dimensional systems has to be calculated by the use of 'defect' particles already explained in \sref{subsec:1_dimensional_systems}.

\section{Conclusion and Outlook}
\label{sec:conclusion}
In this work we have employed the Kato-Bloch perturbation approach to calculate the ground state energies and boundaries of the  Mott-insulator phase, both  for various Bose-Hubbard and the Jaynes-Cummings lattice systems. 

The Kato-Bloch approach leads to explicit formulas for any perturbation order, i.e. there is a clear distinction of the orders in contrast to the mixed expressions in the Schr\"odinger-Rayleigh perturbation theory. Moreover, all the appearing expressions can be represented diagrammatically, which allows to utilize knowledge from graph theory to speed up the numerical algorithms considerably. In fact we were able to perform all calculations on an off-the-shelf computer using Matlab in appropriate time.

Additionally to these advantages, the comparison of our results with those from other methods, including DMRG and VCA prove the high accuracy of our implementation.

Building upon the work by Teichmann et al.~\cite{teichmann_process-chain_2009}, who determined the ground state energy, the boundary of the Mott-insulator phase and various other quantities for the BH system in detail, we focused on adding disorder to the BH model, applying this approach to the JCL model and on solving the problem that appears when trying to calculate the phase boundary for 1-dimensional systems with the method of effective potential.

The Kato-Bloch approach is an extremely powerful method and could be easily generalized to deal with other lattice systems apart from the BH and the JCL as well. Especially interesting in our eyes would be the implementation of models with nearest neighbour interaction or an additional superlattice structure, which exhibit a richer phase diagram as the BH and the JCL model. 

\ack{
The authors want to thank N. Teichmann for providing their process chain approach data to countercheck our results. We also want to thank H. Monien for allowing us to use their DMRG results to compare with ours. Last but not least we want to thank E. Arrigoni and M. Knap for many fruitful discussions on this topic. This work was partly supported by the Austrian Science Fund (FWF  P18551-N16).}

\end{document}